\documentclass[aps,prx,twocolumn,reprint,amssymb,floats,superscriptaddress]{revtex4-2}

\usepackage[title]{appendix}
\usepackage{xr}
\usepackage{graphicx}  
\usepackage{multirow}
\usepackage{svg}
\usepackage{physics}
\usepackage{tikz}

\usepackage[normalem]{ulem}
\usepackage{hyperref}

\usepackage{longtable}
\usepackage[T1]{fontenc}
\usepackage{dcolumn}   

\usepackage{bm}        
\usepackage{amsfonts}  
\usepackage{amsmath}   
\usepackage{amssymb}   
\usetikzlibrary{quantikz}

\pgfdeclarelayer{bg}    
\pgfsetlayers{bg,main}  

\usetikzlibrary{decorations.markings}

\tikzset{
  pics/BoxEnv/.style args={#1,#2,#3,#4}{
     code={
       \draw[fill=blue!20, rounded corners] (0,0) -- (0,#2) -- (#3*#1,#2) -- (#3*#1,0.4) -- ($(0.5*#1,0.4) - (0.5*#3*#1,0)$) -- ($(0.5*#1,#2) - (0.5*#3*#1,0)$) --($(0.5*#1,#2) + (0.5*#3*#1,0)$)--($(0.5*#1,0.4) + (0.5*#3*#1,0)$)--($(#1,0.4) - (#3*#1,0)$)--($(#1,#2) - (#3*#1,0)$)--(1*#1,#2)--(1*#1,0)--cycle;
       \node[] (E) at (0.5*#1,0.2) {#4};
     }
  }
}

\tikzset{
  pics/Tdiag/.style args={#1,#2}{
     code={
     \scalebox{0.8}{
       \node[rounded corners,rectangle,minimum width=2.5em] (A) at (0,0.4) {};
    \node[rounded corners,rectangle,minimum width=2.5em] (B) at (0,-0.4) {};
    \node[rounded corners,rectangle,minimum width=2.5em, minimum height=3.em,draw] (T) at (0,0) {$E$};

    \node[rounded corners,rectangle,minimum height=1.5em,draw] (U) at (-0.8,0) {#1};
    \node[rounded corners,rectangle,minimum height=1.5em,draw] (V) at (0.85,0) {#2};
    \draw [double, rounded corners] (A.west) -| (U.north) node [label=above:{$i_1$}] {};
    \draw [double, rounded corners] (A.east) -| (V.north) node [label=above:{$i_2$}] {};
    \draw [double, rounded corners] (B.west) -| (U.south) node [label=below:{$o_1$}] {};
    \draw [double, rounded corners] (B.east) -| (V.south) node [label=below:{$o_2$}] {};}
     }
  }
}

\def\EBdiag{
\begin{tikzpicture}[baseline=-0.5ex, scale = 0.9] 
\draw[rounded corners=5pt, line width=1pt] (-0.8,-0.5) rectangle (0.8,-0.1);
\draw[rounded corners=5pt, line width=1pt] (-0.8,+0.1) rectangle (0.8, .5);
\draw[rounded corners=4pt, line width=1pt, fill = white] (-0.7+ 0.15 ,-0.3) rectangle (-0.25+ 0.1 ,0.3);
\draw[rounded corners=4pt, line width=1pt, fill = white] (0.25 - 0.1,-0.3) rectangle (0.7 - 0.15,0.3);
\node at (-1 + 0.65,0) {$E$};
\node at (1 - 0.65,0) {$B_i$};
\end{tikzpicture}}

\newcommand{\TdiagEmbel}[2]{\scalebox{0.85}{\begin{tikzpicture}[baseline=-0.5ex,
    ] 
    \node[rounded corners,rectangle,minimum width=2.5em] (A) at (0,0.4) {};
    \node[rounded corners,rectangle,minimum width=2.5em] (B) at (0,-0.4) {};
    \node[rounded corners,rectangle,minimum width=2.5em, minimum height=3.em,draw] (T) at (0,0) {$E$};

    \node[rounded corners,rectangle,minimum height=1.5em,draw] (U) at (-0.8,0) {$#1$};
    \node[rounded corners,rectangle,minimum height=1.5em,draw] (V) at (0.85,0) {$#2^\dagger$};
    \draw [double, rounded corners,postaction={decoration={
    markings, mark=at position 0.5 with {\arrow{stealth}}},decorate}] (A.west) -| (U.north) node [label=above:{$i_1$}] {};
    \draw [double, rounded corners,postaction={decoration={
    markings, mark=at position 0.5 with {\arrow{stealth}}},decorate}] (A.east) -| (V.north) node [label=above:{$i_2$}] {};
    \draw [double, rounded corners, postaction={decoration={
    markings, mark=at position 0.2 with {\arrowreversed{stealth}}},decorate}] (B.west) -| (U.south) node [label=below:{$o_1$}] {};
    \draw [double, rounded corners, postaction={decoration={
    markings, mark=at position 0.2 with {\arrowreversed{stealth}}},decorate}] (B.east) -| (V.south) node [label=below:{$o_2$}] {};
\end{tikzpicture}}}

\newcommand{\Tdiag}[2]{\scalebox{0.85}{\begin{tikzpicture}[baseline=-0.5ex,decoration={
    markings,
    mark=at position 0.5 with {\arrow{stealth}}}
    ] 
    \node[rounded corners,rectangle,minimum width=2.5em] (A) at (0,0.4) {};
    \node[rounded corners,rectangle,minimum width=2.5em] (B) at (0,-0.4) {};
    \node[rounded corners,rectangle,minimum width=2.5em, minimum height=3.em,draw] (T) at (0,0) {$E$};

    \node[rounded corners,rectangle,minimum height=1.5em, draw] (U) at (-0.8,0) {$#1$};
    \node[rounded corners,rectangle,minimum height=1.5em, draw] (V) at (0.85,0) {$#2^\dagger$};
    \draw [double, rounded corners] (A.west) -| (U.north);
    \draw [double, rounded corners] (A.east) -| (V.north);
    \draw [double, rounded corners] (B.west) -| (U.south);
    \draw [double, rounded corners] (B.east) -| (V.south);
\end{tikzpicture}}}

\newcommand{\TdiagWide}[2]{\scalebox{0.85}{\begin{tikzpicture}[baseline=-0.5ex,decoration={
    markings,
    mark=at position 0.5 with {\arrow{stealth}}}
    ] 
    \node[rounded corners,rectangle,minimum width=2.5em] (A) at (0,0.4) {};
    \node[rounded corners,rectangle,minimum width=2.5em] (B) at (0,-0.4) {};
    \node[rounded corners,rectangle,minimum width=2.5em, minimum height=3.em,draw] (T) at (0,0) {$E$};

    \node[rounded corners,rectangle,minimum height=1.5em] (U) at (-1,0) {$#1$};
    \node[rounded corners,rectangle,minimum height=1.5em] (V) at (1.05,0) {$#2^\dagger$};
    \draw [double, rounded corners] (A.west) -| (U.north);
    \draw [double, rounded corners] (A.east) -| (V.north);
    \draw [double, rounded corners] (B.west) -| (U.south);
    \draw [double, rounded corners] (B.east) -| (V.south);
    \node[rounded corners,rectangle,minimum height=1.5em, fill=white, draw] at (-1,0) {$#1$};
    \node[rounded corners,rectangle,minimum height=1.5em, fill=white, draw] at (1.05,0) {$#2^\dagger$};
\end{tikzpicture}}}

\newcommand{\TdiagB}[3]{\scalebox{0.85}{\begin{tikzpicture}[baseline=-0.5ex,decoration={
    markings,
    mark=at position 0.5 with {\arrow{stealth}}}
    ] 
    \node[rounded corners,rectangle,minimum width=2.5em] (A) at (0,0.4) {};
    \node[rounded corners,rectangle,minimum width=2.5em] (B) at (0,-0.4) {};
    \node[rounded corners,rectangle,minimum width=2.5em, minimum height=3.em,draw] (T) at (0,0) {$#1$};

    \node[rounded corners,rectangle,minimum height=1.5em, draw] (U) at (-0.8,0) {$#2$};
    \node[rounded corners,rectangle,minimum height=1.5em, draw] (V) at (0.85,0) {$#3^\dagger$};
    \draw [double, rounded corners] (A.west) -| (U.north);
    \draw [double, rounded corners] (A.east) -| (V.north);
    \draw [double, rounded corners] (B.west) -| (U.south);
    \draw [double, rounded corners] (B.east) -| (V.south);
\end{tikzpicture}}}

\newcommand{\LinearEnvdiag}[2]{\scalebox{0.85}{\begin{tikzpicture}[baseline=-0.5ex,decoration={
    markings,
    mark=at position 0.5 with {\arrow{stealth}}}
    ] 
    \node[rounded corners,rectangle,minimum width=4em] (A) at (0,0.4) {};
    \node[rounded corners,rectangle,minimum width=4em] (B) at (0,-0.4) {};

    \node[rounded corners,rectangle,minimum height=1.5em, draw] (U) at (-1.2,0) {$#1$};
    \node[rounded corners,rectangle,minimum height=1.5em, draw] (V) at (1.2,0) {$#2^\dagger$};
    \draw [double, rounded corners] (A.west) -| (U.north);
    \draw [double, rounded corners] (A.east) -| (V.north);
    \draw [double, rounded corners] (B.west) -| (U.south);
    \draw [double, rounded corners] (B.east) -| (V.south);

    \node[rounded corners, fill=white, rectangle,minimum width=0.5em, minimum height=3.em, draw] (Ed) at (-0.38,0) {$E_{\rm L}^\dagger$};
    \node[rounded corners, fill=white, rectangle,minimum width=0.5em, minimum height=3.em,draw] (E) at (0.38,0) {$E_{\rm L}$};

    \node[rounded corners, red, dashed, rectangle,minimum width=5em, minimum height=4.em,draw] (E) at (0,0) [label={[red]below:{$E_{\rm quad}$}}] {}; 
    
\end{tikzpicture}}}

\newcommand{\Tderiv}[1]{\scalebox{0.85}{\begin{tikzpicture}[baseline=-0.5ex,decoration={
    markings,
    mark=at position 0.5 with {\arrow{stealth}}}
    ] 
    \node[rounded corners,rectangle,minimum width=2.5em] (A) at (0,0.4) {};
    \node[rounded corners,rectangle,minimum width=2.5em] (B) at (0,-0.4) {};
    \node[rounded corners,rectangle,minimum width=2.5em, minimum height=3.em,draw] (T) at (0,0) {$E$};

    \node[rounded corners,rectangle,minimum height=1.5em] (U) at (-0.8,0) {};
    \node[rounded corners,rectangle,minimum height=1.5em, draw] (V) at (0.85,0) {$#1^\dagger$};
    \draw [double distance =2pt, rounded corners] (A.west) -| (U.north);
    \draw [double distance =2pt, rounded corners] (A.east) -| (V.north);
    \draw [double distance =2pt, rounded corners] (B.west) -| (U.south);
    \draw [double distance =2pt, rounded corners] (B.east) -| (V.south);
\end{tikzpicture}}}

\newcommand{\Tderivderiv}[1]{\scalebox{0.85}{\begin{tikzpicture}[baseline=-0.5ex,decoration={
    markings,
    mark=at position 0.5 with {\arrow{stealth}}}
    ] 
    \node[rounded corners,rectangle,minimum width=2.5em] (A) at (0,0.4) {};
    \node[rounded corners,rectangle,minimum width=2.5em] (B) at (0,-0.4) {};
    \node[rounded corners,rectangle,minimum width=2.5em, minimum height=3.em,draw] (T) at (0,0) {$#1$};

    \node[rounded corners,rectangle,minimum height=1.5em] (U) at (-0.8,0) {};
    \node[rounded corners,rectangle,minimum height=1.5em] (V) at (0.8,0) {};
    \draw [double distance =2pt, rounded corners] (A.west) -| (U.north);
    \draw [double distance =2pt, rounded corners] (A.east) -| (V.north);
    \draw [double distance =2pt, rounded corners] (B.west) -| (U.south);
    \draw [double distance =2pt, rounded corners] (B.east) -| (V.south);
\end{tikzpicture}}}

\newcommand{\ABUVdiag}[4]{
\scalebox{0.85}{
\begin{tikzpicture}[baseline=0ex]
    \node[rounded corners,rectangle,minimum width=2.5em,draw] (A) at (0,0.4) {$#1$};
    \node[rounded corners,rectangle,minimum width=2.5em,draw] (B) at (0,-0.4) {$#2$};
    \node[rounded corners,rectangle,minimum height=1.5em,draw] (U) at (-0.8,0) {$#3$};
    \node[rounded corners,rectangle,minimum height=1.5em,draw] (V) at (0.85,0) {$#4^\dagger$};
    \draw [thick, rounded corners] (A.west) -| (U.north);
    \draw [thick, rounded corners] (A.east) -| (V.north);
    \draw [thick, rounded corners] (B.west) -| (U.south);
    \draw [thick, rounded corners] (B.east) -| (V.south);
\end{tikzpicture}}}

\newcommand{\ABUVdiagClean}[2]{
\scalebox{0.85}{
\begin{tikzpicture}[baseline=0ex]
    \node[rounded corners,rectangle,minimum width=2.5em,draw] (A) at (0,0.4) {$#1$};
    \node[rounded corners,rectangle,minimum width=2.5em,draw] (B) at (0,-0.4) {$#2$};
    \node[rounded corners,rectangle,minimum height=1.5em] (U) at (-0.8,0) {};
    \node[rounded corners,rectangle,minimum height=1.5em] (V) at (0.85,0) {};
    \draw [thick, rounded corners] (A.west) -| (U.north);
    \draw [thick, rounded corners] (A.east) -| (V.north);
    \draw [thick, rounded corners] (B.west) -| (U.south);
    \draw [thick, rounded corners] (B.east) -| (V.south);
\end{tikzpicture}}}

\newcommand{\PiPj}[4]{
\scalebox{0.85}{
\begin{tikzpicture}[baseline=0ex]
    \node[rounded corners,rectangle,minimum width=2.5em,draw] (A) at (0,0.4) {$#1$};
    \node[rounded corners,rectangle,minimum width=2.5em,draw] (B) at (0,-0.4) {$#2$};
    
    \node[rounded corners,rectangle,minimum width=2.5em,draw] (U) at (0,1.) {$#3$};
    \node[rounded corners,rectangle,minimum width=2.5em,draw] (V) at (0,-1.07)  {$#4$};
    \draw [thick, rounded corners] (A.west) -| (-0.7,0.8) |- (U.west);
    \draw [thick, rounded corners] (A.east) -| (+0.7,0.8) |- (U.east);
    \draw [thick, rounded corners] (B.west) -| (-0.7,-0.8) |- (V.west);
    \draw [thick, rounded corners] (B.east) -| (+0.7,-0.8) |- (V.east);
\end{tikzpicture}}}

\newcommand{\UVSpiderDiagram}{
    \begin{tikzpicture}[baseline=-0.8ex, trim left=-2.5em, trim right=+2.5em]]
        \node[rounded corners,rectangle,minimum width=2.5em] (A1) at (-1.1,-0.5) {};
        \node[rounded corners,rectangle,minimum width=2.5em] (B1) at (-1.1,+0.5) {};

        \node[rounded corners,rectangle,minimum width=2.5em] (A2) at (+1.1,-0.5) {};
        \node[rounded corners,rectangle,minimum width=2.5em] (B2) at (+1.1,+0.5) {};
        
        \node[rectangle, rounded corners, minimum height=1.5em, draw] (U) at (-0.4, 0) {$U$};
        \draw [thick, rounded corners] (B1.east) -| (U.north);
        \draw [thick, rounded corners] (A1.east) -| (U.south);

        \node[rectangle, rounded corners, minimum height=1.5em, draw] (V) at (0.4, 0) {$U^\dagger$};
        \draw [thick, rounded corners] (B2.west) -| (V.north);
        \draw [thick, rounded corners] (A2.west) -| (V.south);
    \end{tikzpicture}
}

\newcommand{\UVSpiderDiagramAnnotated}[2]{
    \begin{tikzpicture}[baseline=-0.8ex, trim left=-2.5em, trim right=+2.5em]]
        \node[rounded corners,rectangle,minimum width=2.5em] (A1) at (-1.1,-0.5) {};
        \node[rounded corners,rectangle,minimum width=2.5em] (B1) at (-1.1,+0.5) {};

        \node[rounded corners,rectangle,minimum width=2.5em] (A2) at (+1.1,-0.5) {};
        \node[rounded corners,rectangle,minimum width=2.5em] (B2) at (+1.1,+0.5) {};
        
        \node[rectangle, rounded corners, minimum height=1.5em, draw] (U) at (-0.4, 0) {$#1$};
        \draw [thick, rounded corners] (B1.east) -| (U.north);
        \draw [thick, rounded corners] (A1.east) -| (U.south);

        \node[rectangle, rounded corners, minimum height=1.5em, draw] (V) at (0.4, 0) {$#2$};
        \draw [thick, rounded corners] (B2.west) -| (V.north);
        \draw [thick, rounded corners] (A2.west) -| (V.south);
    \end{tikzpicture}
}

\newcommand{\UVSpiderDiagramWide}[2]{
    \begin{tikzpicture}[baseline=-0.8ex, trim left=-2.5em, trim right=+2.5em]]
        \node[rounded corners,rectangle,minimum width=2.5em] (A1) at (-1.25,-0.5) {};
        \node[rounded corners,rectangle,minimum width=2.5em] (B1) at (-1.2,+0.5) {};

        \node[rounded corners,rectangle,minimum width=2.5em] (A2) at (+1.25,-0.5) {};
        \node[rounded corners,rectangle,minimum width=2.5em] (B2) at (+1.2,+0.5) {};
        
        \node[rectangle, rounded corners, minimum height=1.5em, draw] (U) at (-0.65, 0) {$#1$};
        \draw [thick, rounded corners] (B1.east) -| (U.north);
        \draw [thick, rounded corners] (A1.east) -| (U.south);

        \node[rectangle, rounded corners, minimum height=1.5em, draw] (V) at (0.65, 0) {$#2$};
        \draw [thick, rounded corners] (B2.west) -| (V.north);
        \draw [thick, rounded corners] (A2.west) -| (V.south);
    \end{tikzpicture}
}

\newcommand{\ParallelLines}[2]{
\begin{tikzpicture}[baseline=-0.6ex]
  \tikzstyle{line}=[-, thick]
  
  \draw[line] (0, #1) -- (0.5, #1);
  
  \draw[line] (0, #2) -- (0.5, #2);
\end{tikzpicture}}

\pgfdeclarelayer{nodelayer}
\pgfdeclarelayer{edgelayer}
\pgfsetlayers{edgelayer,nodelayer,main}
\tikzstyle{new style 0}=[fill=white, draw=black, shape=rectangle]
\tikzstyle{medium box}=[fill=white, draw=black, shape=rectangle, rounded corners, minimum size = 5]

\newcommand{\re}[1]{\mathrm{Re}\left\{#1\right\}}

\newcommand{\pwisein}{\left\{ \begin{array}{ll}}
\newcommand{\pwiseout}{\end{array}\right.}

\newcommand{\kket}[1]{\| #1 \rangle\!\rangle}
\newcommand{\bbra}[1]{ \langle\!\langle#1\| }
\DeclareMathOperator*{\argmin}{arg\,min}
\DeclareMathOperator*{\E}{\mathbb{E}}

\usepackage{mathtools}

\tikzset{
  pics/BoxEnv/.style args={#1,#2,#3,#4}{
     code={
       \draw[fill=blue!20, rounded corners] (0,0) -- (0,#2) -- (#3*#1,#2) -- (#3*#1,0.4) -- ($(0.5*#1,0.4) - (0.5*#3*#1,0)$) -- ($(0.5*#1,#2) - (0.5*#3*#1,0)$) --($(0.5*#1,#2) + (0.5*#3*#1,0)$)--($(0.5*#1,0.4) + (0.5*#3*#1,0)$)--($(#1,0.4) - (#3*#1,0)$)--($(#1,#2) - (#3*#1,0)$)--(1*#1,#2)--(1*#1,0)--cycle;
       \node[] (E) at (0.5*#1,0.2) {#4};
     }
  }
}

\begin{document}

\title{Quantum landscape tomography for efficient single-gate optimization on quantum computers}
\author{Matan Ben Dov}
\affiliation{Department of Physics, Bar-Ilan University, 52900 Ramat Gan, Israel}
\affiliation{Center for Quantum Entanglement Science and Technology, Bar-Ilan University, 52900 Ramat Gan, Israel}
\author{Itai Arad}
\affiliation{Centre for Quantum Technologies, National University of Singapore, 117543 Singapore, Singapore}
\author{Emanuele G Dalla Torre}
\affiliation{Department of Physics, Bar-Ilan University, 52900 Ramat Gan, Israel}
\affiliation{Center for Quantum Entanglement Science and Technology, Bar-Ilan University, 52900 Ramat Gan, Israel}

\date{\today}

\begin{abstract}  
Circuit optimization is a fundamental task for practical applications of near-term quantum computers. In this work we address this challenge through the powerful lenses of tensor network theory. Our approach involves the full characterization of the influence of individual gates on the entire circuit, a process we call quantum landscape tomography. We derive the necessary and sufficient requirements of this process and propose two implementations, respectively based on 2-unitary design and Clifford tableaux. The latter implementation strikes a convenient balance between the number of shots and the number of circuits needed for the tomography. Numerical simulations based on a realistic noise model demonstrate the advantage of our approach with respect to both gradient-free and gradient-based methods. Overall, our findings highlight the potential of quantum landscape tomography to enhance circuit optimization in near-term quantum computing applications.

\end{abstract}


\maketitle 


\section{Introduction}

\begin{figure}[t]
    \centering
    \includegraphics[width=\columnwidth]{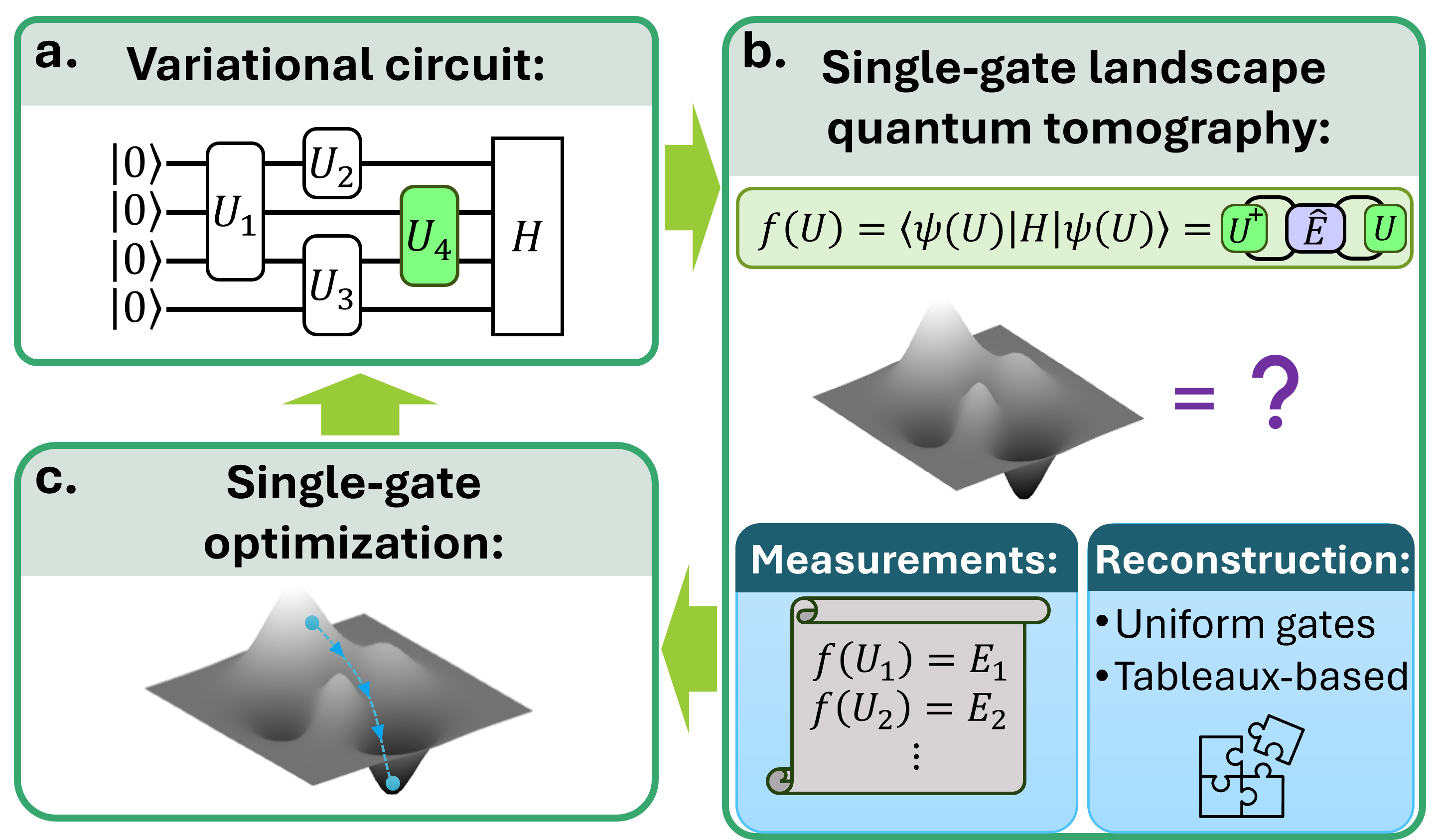}
    \caption{Schematic diagram of the optimization algorithm for dense variational quantum circuits proposed in this paper. {\bf a.} The algorithm starts by picking a single $k$-qubit gate from the variational circuit, for example $U_4$. {\bf b.} We obtain a full description of the cost function dependency on the chosen gate $f(U_4)$ by performing {\it quantum landscape tomography}, an algorithm that estimates the full cost function from a sequence of gate substitutions and measurements. {\bf c.} We find the optimal gate using classical optimization of the reconstructed cost function, and replace the previous gate with the optimal one. The algorithm is repeated many times, iterating over all gates, until convergence is reached.}
    \label{fig:main_diag}
\end{figure}

In recent years, substantial efforts have been devoted to developing quantum algorithms fitted to noisy intermediate-scale quantum (NISQ) computers \cite{preskill2018quantum, bharti2022noisy,cerezo2021variational,huang2023near}. In this quest, variational quantum algorithms are one of the main candidates with the potential to distil special properties of noisy devices through hybrid algorithms that combine classical and quantum calculations \cite{perdomo2018opportunities, endo2021hybrid, callison2022hybrid}. 
Consequently the efficient optimization of variational quantum circuits has substantial importance for the effectiveness of quantum algorithms in the NISQ era.



Current methods for quantum circuit optimization can be generically divided in two main types: gradient-based \cite{verdon2018universal, crooks2019gradients,schuld2019evaluating,sweke2020stochastic,hubregtsen2022single}, and gradient free \cite{lavrijsen2020classical, wiedmann2023empirical,bonet2023performance}. In the gradient-based approaches the optimization is guided by the gradients of the cost function, which are separately measured on the quantum hardware. 
These methods are a natural extension of classical backpropagation used in classical  machine learning optimization \cite{rumelhart1986learning,lecun1989backpropagation,lillicrap2020backpropagation}. Measuring the gradients on quantum hardware presents its own challenges, as using simple infinitesimal differentiation by small perturbation of the parameters significantly amplifies the sampling noise. For this reason,  gradients are commonly measured using the parameter-shift rule algorithm \cite{mitarai2018quantum,schuld2019evaluating,crooks2019gradients,izmaylov2021analytic, wierichs2022general, kyriienko2021generalized}, 
based on the observation that  the parameters of a quantum circuit are often rotation angles periodic in $2\pi$. In this case, the gradients can be measured using finite differentiation steps, by shifting the parameters by a discrete shift of the rotation angles, typically $\pi/2$. This technique can be further expanded by considering higher-order derivatives and natural gradients, or minimizing the number of measurements needed to extract these quantities \cite{mari2021estimating,rebentrost2019quantum,teo2023optimized, wiersema2023here,wiersema2023optimizing}. 
Unfortunately, unlike backpropagation in classical machine learning, the measurement of gradients remains demanding in terms of quantum resources: the number of measurements required for measuring the full gradient grows linearly with the number of variational parameters, which is usually large for practical applications. 
In addition, the whole measurements needs to be repeated after each infinitesimal improvement of the cost function, making this approach inefficient. 

The second type of optimization algorithms, gradient-free approaches, ``gives up'' on measuring the gradients and instead dedicates all its resources to evaluate the function at different points of the parameter space. Common gradient-free optimizers such as SPSA \cite{spall1992multivariate, wiedmann2023empirical} and COBYLA \cite{powell2007view,lavrijsen2020classical, miki2022variational} treat the cost function as a ``black-box'' and deploy different search strategies to find global minima. In practice, gradient-free optimizers were found to be more convenient, especially in noisy environments 
\cite{sung2020using, singh2023benchmarking}. Nevertheless, such algorithms usually require large number of shots in order to reach good convergence.



In this work we present a different optimization algorithm for parametric quantum circuits. We adopt a coordinate descent approach, performing a full optimization on only a subset of the parameters; see Refs.~\cite{nesterov2012efficiency,wright2015coordinate,shi2016primer} for related approaches in classical optimization and machine learning tasks. 
This approach is at the core of the Evenbly-Vidal method for tensor network optimization \cite{evenbly2015tensor, hauru2021riemannian}, where one sequentially replaces each tensor with its optimal counterpart.
%
Simple implementations of this approach to quantum circuits have been recently introduced by Refs.~\cite{vidal2018calculus, parrish2019jacobi, nakanishi2020sequential, ostaszewski2021structure} for a single rotation parameter, 
and by Ref.~\cite{wada2024sequential} for generic single-qubit gates.
Here, we extend this approach to generic $k$-qubit gates and demonstrate its application to the case of 2-qubit gates.

Our approach strives for an optimal balance between a low count of required quantum circuits and a small overhead in the number of shots.
Instead of measuring local gradients of the cost function with respect to all possible directions, we propose to characterize the full dependency of the cost function on the parameters that characterize a single multi-qubit gate. 
At each iteration we pick a single $k$-qubit gate from the quantum circuit and perform a full {\it quantum landscape tomography} (also called environment tensor tomography) on that gate. 
This step allows us to recover the full dependency of the cost function on a single $k$-qubit gate by evaluating the cost function with different gates substitutions while keeping the other gates fixed. The next step is then to search for the optimal gate giving rise to the lowest cost function value. This step can be performed efficiently using classical numerical optimization. Finally we replace the gate with the optimal solution and move to the next one, starting over the tomography and optimization subroutines. 

Our coordinate descent approach has three key advantages with respect to gradient-based and gradient-free optimization.
First, our method features an efficient allocation of circuit evaluations compared to the parameter-shift rule: the number of unique circuits required for a single gate tomography does not scale up with the size of the system. On the other hand, our approach allow us to take large steps, extracting more value from previous measurements, with respect to gradient descent approaches. Finally, our approach relies on the analytical structure of the cost function and is therefore advantageous over black-box optimization schemes. 

The paper is structured as follows: in Sec.~\ref{sec:preliminaries} we define the local cost function and the environment tensor used throughout this paper and discuss the preconditions for performing a successful tomography of the environment tensor. In Sec.~\ref{sec:env_tomo} we describe a general framework for environment tensor tomography and implement it for two choices of gate sets, random unitary gates, and a tablueax-based construction of Clifford gates. In Sec.~\ref{sec:minimum}, we explain how to use environment tomography in a circuit optimization algoritm. In Sec.~\ref{sec:numerical} we implement our algorithm on simulated quantum systems and compare it with gradient-based and gradient-free optimizers. In Sec.~\ref{sec:compare} we discuss some additional features and potential extensions of the technique presented in this work, as well as compare it with the parameter-shift rule formula and its variants. Section \ref{sec:conclude} summarizes the paper and suggests further extensions to the techniques introduced in this work.









\section{Preliminaries\label{sec:preliminaries}}

\subsection{Optimization methods for VQE}


The problem of optimizing a variational circuits is relevant to a large variety of quantum algorithms, including
variational quantum algorithms (VQA) \cite{endo2021hybrid,cerezo2021variational} such as variational quantum eigensolvers (VQE) for finding ground states of molecular Hamiltonians \cite{peruzzo2014variational,fedorov2022vqe,tilly2022variational}, quantum machine learning (QML) algorithms such as variational quantum classifiers \cite{mitarai2018quantum,schuld2020circuit}, autoencoders \cite{romero2017quantum,cao2021noise}, quantum circuit recompilation 
\cite{khatri2019quantum,heya2018variational} and quantum state encoding \cite{bendov2024approximate,ran2020encoding,rudolph2023decomposition}. 

The cost function of a generic variational algorithm has the form of
\begin{align}
    C(\theta) = \sum_{i}{{\rm Tr}\left(  \rho_i U^\dagger(\theta)\hat{H}_i U(\theta)\right)},
    \label{eq:Ctheta}
\end{align}
which consist of three main ingredients: an initial state $\rho_i$, a family of unitary circuits or {\it ansatz}, $U(\theta)$, and a measured Hermitian operator $H_i$. The summation over $i$ can be used to average over a set of initial states and operators, which can be useful in QML algorithms. The ansatz $U(\theta)$ determines the space of circuits explored during the optimization
and needs to strike a balance between expressibility - the ability to implement a large variety of unitary evolutions, and trainability - the ability to undergo optimization and converge to a minimum using a manageable number of iterations \cite{holmes2022connecting,nakaji2021expressibility,du2022efficient}.

For concreteness, we focus on the specific case of VQE. Its cost function is the expectation value of a many-body Hamiltonian in the encoded quantum state:
\begin{equation}
    C_{\rm VQE}(\{U_i\}) = \Tr{\hat{H} U_{\rm circ}\ketbra{0}{0} U_{\rm circ}^\dagger}
    \label{eq:CVQE}
\end{equation}
Practically, $H$ is often decomposed into a sum of Pauli strings. The expectation value of each Pauli string is obtained by measuring each qubit in the appropriate Pauli basis. The contributions of all Pauli strings are then summed up, giving rise to an expression analogous to Eq.(\ref{eq:Ctheta}). 

The circuit $U_{\rm circ}$ corresponds to a sequence of local unitary gates,  $U_{\rm circ}= U_{N_{\rm g}}U_{N_{\rm g}-1}...U_2 U_1$, where $\{U_i\}_{i=1..N_{\rm g}}$ are usually 1- or 2-qubit gates and will later be referred to as $k$-qubit gates. In this work, we choose as a baseline a ``densely parameterized'' ansatz inspired by tensor networks calculations, in which each gate is fully tunable, but their sequence is fixed. This approach allows us to use insights from tensor networks schemes to better analyze the properties of the cost function. We will show that our method of optimization can be adapted to more restrictive circuits by constraining the properties of the gates during gradient measurements (see Section \ref{sec:constr}).



\begin{figure}
\centering
\includegraphics[width=\columnwidth]{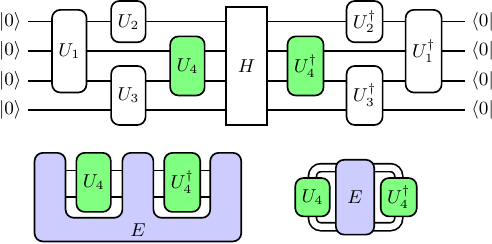}
\caption{ Three equivalent notations for the tensor network used to calculate the expectation value of $H$ in the encode state $U_{\rm circ}|0\rangle$, see Eq. (\ref{eq:CVQE}), highlighting the role of a single 2-qubit gate $U_4$ }
\label{fig:EnvTens}
\end{figure}

To introduce our method, we first examine the dependency of the cost function on a single $k$-qubit gate. For that end, we can view the cost function from the perspective of tensor calculus, treating each $k$-qubit gate as a tensor with k input and k output indices. 
In this picture, the quantum state at the end of the circuit is obtained by contracting the tensor network of the different gate tensors, starting from the initial zero state, and the final expectation value of the Hamlitonian is then calculated by contracting the tensor of final state from both sides of the Hamiltonian operator (see Fig.~\ref{fig:EnvTens}). 
This notation helps the contraction process used for calculating the cost function because the contractions can be performed out-of-order, in contrast to the simulation of the unitary evolution of a quantum state, where time ordering must be enforced. On the other hand, for the same reason, it is less natural to account for the unitarity of quantum gates, which needs to be required separately.

Through the lenses of tensor calculus, the dependence
of the cost function on a single $k$-qubit gate
can be described by an environment tensor $E$, obtained by contracting all the tensor network but the two copies of the chosen gate, as presented in Fig. \ref{fig:EnvTens}. 
This contraction simplifies the cost function to the following reduced bi-linear form
\begin{align}
f_{\rm exact}(U) &\equiv C_{\rm VQE}(U_j = U)  \\
   &=\sum_{{i_1},{i_2},{o_1},{o_2}}{{U}_{\Vec{i_1}}^{\Vec{o_1}} E^{{i_1},{i_2}}_{{o_1},{o_2}} \left({U}_{\Vec{i_2}}^{\Vec{o_2}}\right)^*} 
   \\&= \TdiagEmbel{U}{U}\label{eq:reduced_cost_function}
\end{align}
where $i_{1/2}$, $o_{1/2}$ respectively denote input and output indices of the tensor $E$. The reduced cost function $f_{\rm exact}(U)$ is a real-valued bi-linear function of a single unitary gate, $U$. 
The environment tensor can be interpreted in different ways, such as a non-positive inner product on the set of unitary gates, or as a non-convex sum of quantum channels. More properties of the environment tensor and the reduced cost function are detailed in Appendix \ref{app:environment_tensor}.


The cost function $f_{\rm exact}$ is the expectation value of the energy of the state obtained by averaging over infinitely many shots.
On quantum hardware, however, expectation values are evaluated using a finite number of shots, each giving a single noisy evaluation of the energy of the state. This variance in energy for different shots is important for estimating the reconstruction accuracy of the environment tensor, and it is therefore important to keep track of the stochastic part of the function as well. We denote the random variable of single-shot energy value as $f_\text{single-shot}(U)$, whose mean value gives back $f_{\rm exact}(U)$
\begin{align}
    f_{\rm exact}(U) = \mathbb{E}[f_\text{single-shot}(U)].
\end{align}
In what follows, we discuss how to estimate the environment tensor from a finite number of single-shot measurements $f_\text{single-shot}(U)$.

\subsection{Probing the bi-linear environment tensor \label{ch:env_tens}}



{Environment tomography shares some similarities with different methods of channel tomography \cite{kunjummen2023shadow, levy2021classical, mohseni2008quantum}, and in particular, with ancilla-assisted process tomography \cite{scott2008optimizing, xue2022variational}. These algorithms make use of unitary gate substitutions in order to extract information about the channel or environment tensor, while holding distinct features. Process tomography is more flexibile in the application of unitaries, and some scheme make use of different input and output unitary operators, whereas in environment tensor tomography we are restricted to the structure of the variational circuit. Additionally, unlike in channel tomography, the environment tensor does not follow the complete-positivity and trace preserving properties.}

A naive way to perform the tomography of an environment tensor $E$ consists of applying two different operators, $U$ and $V$, on its left and right indexes, giving rise to $f(U,V)$, obtained by substituting $U^\dagger$ with $V$ in Eq. \ref{eq:reduced_cost_function}. For example, by probing $f(P_i,P_j)$ with all pairs of Pauli strings $P_i$ and $P_j$ one can map all the components of the environment tensor. Unfortunately, on real quantum hardware, on can only probe the symmetric cost function $f_{\rm exact}(U)=f(U,U^\dagger)$, as $U$ and $U^\dagger$ originate from the physical application of a single gate, which is included twice in the expectation value of the Hamiltonian (once for the bra and once for the kat).

This observation rises the fundamental question: is it possible to perform the full tomography of $E$ by measuring $f(U,U^\dagger)$ with unitary gates $U$ only? In general, this is not the case: as shown in Appendix \ref{app:decomp}, the environment $E$ includes some components that cannot be probed by any symmetric and unitary setup. The good news are that, by definition, the components that cannot be probed do not affect the cost function and, hence, are irrelevant to the optimization protocol.  We refer to the accessible components as {\it relevant} and compute their number in Appendix \ref{app:decomp}. For a generic $k$-qubit environment tensor the number of relevant elements is
\begin{align}
    N_{\rm relevant} = (4^k - 1)^2 + 1 . 
    \label{eq:relevant_dof}
\end{align} 
Clearly $N_{\rm relevant}<D$, with $D= 2^{4k}$, which is the total dimension of a generic rank-4 tensor with an index dimension of $2^k$.
In some special cases the number of degrees of freedom can be reduced, simplifying the tomography process. For example, as shown in Appendix \ref{app:Linear environment tomography}, if the environment tensor can be expressed as the a square of a linear tensor form, the number of degrees of freedom is reduced significantly.

As we will show below, the problem of environment tensor reconstruction can be formalized as a linear regression problem with $N_{\rm relevant}$ degrees of freedom.
In principle, almost any set of $N_{\rm relevant}$ random gates will be sufficient to reconstruct the environment tensor: randomly picked unitaries probe linearly independent parts of the environment tensor, and by inverting a transformation matrix it is possible to extract all the measurable components.
However, in practice, the particular choice of matrices and the reconstruction algorithm influences the susceptibility of the tomography process to stochastic errors, such as those due to the finiteness of the number of measurements (shot-noise). We will demonstrate that reconstructing the environment tensor using random matrices is inefficient, as sub-optimal overlap of a finite number of randomly chosen matrices makes this algorithm more susceptible to noise.

A common alternative to random gates is offered by {\it unitary \textsl{2}-design}, i.e. finite sets of gates whose average for second-order polynomials equals to the average over Haar-random unitary gates. In Appendix \ref{app:opti_2_design} we show that in the absence of information about the environment tensor, a gate set is indeed optimal if and only if it forms a  unitary 2-design.
This approach has several drawbacks: First, the size of unitary 2-design is often much larger than $N_{\rm relevant}$. Even in the simple case of 1-qubit gate, the smallest size of 2-unitary-design sets is 12, which is larger than $N_{\rm relevant}(k=1)=10$ \cite{roy2009unitary}. For 2-qubit gates, there are two common examples of unitary 2-design: the 2-qubit Clifford group \cite{webb2015clifford}, which contains $11520$ gates, and a subgroup of the Clifford group containing $1920$ gates \cite{gross2007evenly}, which is the smallest known group design of 2-qubit gates. Both groups are much larger than the number of relevant components $N_{\rm relevant} = 226$. 
The second drawback of unitary 2-design is that the physical implementation of a specific gate may be more expensive than others in terms of computational time and noise. These considerations are neglected in the above-mentioned proof, which assumes that arbitrary gates can be introduced without any overhead. 


In the next section, we will propose two different variants of the tomography process and analyze their performance for a varying number of shots. Both methods, uniform environment tomography and tableaux-based tomography, use a linear inversion technique to reconstruct the environment tensor from a set of unitary circuit measurements using linear regression.


\section{Environment tomography techniques\label{sec:env_tomo}}

\subsection{Linear-inversion based tomography \label{sec:regression}}





In this section we provide a formal description of environment tensor tomography and estimate the reconstruction accuracy. Let us define a general orthonormal basis that spans the linear space of environment tensors $B = \{B_i\}_{i=1}^{N_b}$, containing $N_b$ elements.  (We refer the reader to Appendix \ref{app:reg} for a basis-free formulation.) The different gates and tensors can be represented in the new coordinate system of $B$. 
We define $\vec{v}_{\rm \hat{E}}$ as the vector of amplitudes of the environment tensor $E$, whose elements are given by the contraction between the tensors $B_i$ and $E$,
\begin{align}
    {\left({v}_{\rm \hat{E}}\right)}_i = \hat{E} \circ B_i  = \EBdiag,
    %
\end{align}
where the operator $\circ$ denotes a full contraction between two tensors.
Using the same basis, one can represent any unitary gate as a vector $\vec{v}_U$, obtained by contracting the projection operator $\pi_{U} = U \otimes U^{\dagger}$ with the basis elements,
\begin{align}
    \left(\vec{v}_{U}\right)_i = \TdiagB{B_i}{U}{U}.
\end{align}
In this coordinate representation, the reduced cost function given by Eq.~(\ref{eq:reduced_cost_function}) can be written as a vector product,
\begin{align}
    f_{\rm exact}(U)=\vec{v}_{U}^T \cdot \vec{v}_{E} = \sum_i\left[\EBdiag  \TdiagB{B_i}{U}{U} \right]. 
\end{align}

This notation offers a useful tool to describe the process of environment tomography. In this process, we ideally evaluate the cost function $f_{\rm exact}(U)$ for a set of $N$ unitary gates $\mathcal{U}_s = \left\{U_1,U_1,...U_{N}\right\}$. If we introduce a matrix $\hat{M}$ that stores rows of projectors $\pi_{U_i}$,
\begin{align}
     \hat{M}_{i,j} = \TdiagB{B_j}{U_i}{U_i},
\end{align}
we can express the relation between $\vec{v}_{E}$ and $f_{\rm exact}$ as
\begin{align}
\hat{M}\cdot \vec{v}_{E} =  \begin{pmatrix}
  f_{\rm exact}(U_1) \\
  f_{\rm exact}(U_2) \\
  \vdots \\
  f_{\rm exact}(U_{N})
  \end{pmatrix}.
  \label{eq:Mv}
  \end{align}



{In reality, the cost function $f$ is estimated each time using a finite number of shots. For the sake of simplicity, we treat each evaluation as a single shot measurement, repeating the same unitary to represent multi-shot evaluations \footnote{For cost function evaluations that require several circuit measurements in different bases, such as VQE with complicated Hamiltonians, each single-shot shadow is evaluated in a single Pauli basis randomly sampled from the different measurement bases. The measurement results are then scaled up to match the Hamiltonian expectation value after averaging over many single-shot measurements.}. We denote $\phi_s$ as a vector that lists all single-shot measurement outcomes
\begin{align}
    \vec{\phi}_s = \begin{pmatrix}
  f_\text{single-shot}(U_1) \\
  f_\text{single-shot}(U_2) \\
  \vdots \\
  f_\text{single-shot}(U_{N_{\rm shots}})
  \end{pmatrix}
\end{align}
where the number of gates $N$ is equal to the number of shots $N = N_{\rm shots}$.
}
{We then define the vector of errors $\vec{\epsilon}$ as the difference between the exact result and single-shot one and, using Eq.~(\ref{eq:Mv}), express it as
\begin{align}
    \vec{\epsilon} = \hat{M} \cdot \vec{v}_{\rm \hat{E}} - \vec{\phi_s}
    \label{eq:linear_reg}
\end{align}
In this work, we focus on the case where $\epsilon$ is due to the finitness of the number of measurement, i.e. shot noise, and we assume $\epsilon$} to be unbiased, i.e., to average to zero with repeating experiments, $\mathbb{E}[\epsilon] = 0 $. 
Performing an environment tomography consists of finding $\vec{v}_{\rm \hat{E}}$ with the smallest possible error, which reduces the tomography to a linear inversion problem. 

The problem at hand is similar to the linear inversion approach used in the context of quantum state tomography \cite{qi2013quantum, nguyen2022optimizing}. The main difference is that in the present case $M$ cannot be diagonal, due to the unavoidable overlaps associated with the physical limitation to  symmetric and unitary gates. 
The representation of the environment tensor using overlapping projections can be formulated using the formal language of {\it frames} in linear algebra, which are vector representations in overcomplete bases \cite{daubechies1986painless, christensen2003introduction, antoine2012frames, innocenti2023shadow}. This similarity allows us to use fundamental results from linear algebra theory, such as the tight frame condition, which translate into a condition to optimal gate ensemble for minimal tomography inaccuracy, as detailed in Appendix \ref{app:frame}.

A specific method to extract $\vec{v}_{\rm \hat{E}}$ is {\it linear regression}, which fits the noisy data by minimizing the total mean squared error (MSE) $\mathbb{E}[\epsilon^2] $, or 
\begin{align}
    \vec{v}_{ \rm op} = \argmin_{\vec{v}_{\rm \hat{E}}} \left\{\norm{\hat{M} \cdot\vec{v}_{\rm \hat{E}} - \vec{\phi}_s}_2^2\right\}
    \label{eq:optimization}
\end{align}
To solve this optimization problem we equate to zero the derivative of the MSE with respect to $\vec{v}_E$, obtaining the linear equation
\begin{align}
    \hat{M}^\dagger \hat{M}  \vec{v} = \hat{M^\dagger} \Vec{\phi}_s
    \label{eq:MtMa}
\end{align}

which by inversion \footnote{For a basis including non-measurable elements, the Moore–Penrose pseudo inverse \cite{ben2003generalized} is used to invert only the non-zero singular values of the second-moment matrix}, gives an estimate for the amplitudes vector:
\begin{align}
    \vec{v}_{ \rm op} = (\hat{M}^\dagger \hat{M})^{-1} \hat{M^\dagger} \Vec{\phi}_s. \label{eq:inv_solution}
\end{align}.

Before we continue, we take a look at the explicit expressions for the two sides of equation \ref{eq:MtMa} that offer valuable insight into the mathematical factors important for performing accurate landscape tomography.
On the left-hand side, $\hat{M}^\dagger \hat{M}$ is a square matrix of size $N_b\times N_b$, which can be interpreted as the second moment, or covariance matrix of the base elements
\begin{align}
    (M^\dagger M)_{i,j} &= N_{\rm shots} \times \underset{U\in \mathcal{U}_s}{\mathbb{E}}\left[B_i^* B_j\right]\\
    &= \sum_{U\in\mathcal{U}_s}{\TdiagB{B_i^*}{U}{U}\;\TdiagB{B_j}{U}{U}} \label{eq:covar_mat}.
\end{align}
This matrix is independent of the measurements outcomes, and represent the coverage of different basis elements by the set of unitary gates, giving larger weight to basis elements which have strongly correlated overlap with the gates.

The right-hand side of Eq.~(\ref{eq:MtMa}), $\hat{M^\dagger} \Vec{\phi}_s$ is a column vector of size $N_b$, which can be interpreted as a weighted sum of all unitary projections $\pi_U$, projected onto the different environment basis elements,
\begin{align}
    M^\dagger \vec{\phi}_s &= \begin{pmatrix}
  \sum\limits_{U\in \mathcal{U}_s} \TdiagB{B_1}{U}{U} \; f_\text{single-shot}(U)\\
  \sum\limits_{U\in \mathcal{U}_s} \TdiagB{B_2}{U}{U} \; f_\text{single-shot}(U) \\
  \vdots \\
  \sum\limits_{U\in \mathcal{U}_s} \TdiagB{B_{N_b}}{U}{U} \; f_\text{single-shot}(U) 
  \end{pmatrix}\\
  &= \sum\limits_{U\in \mathcal{U}_s} 
  f_\text{single-shot}(U) \UVSpiderDiagram \circ \vec{{\rm \mathbf{B}}}, 
  \label{eq:reg_shadow}
\end{align}
where $\vec{{\rm \mathbf{B}}}$ is the column vector that lists all basis elements. We now define the single-shot {\it shadow} 
\begin{align}
    \mathcal{S}_\text{single}(U) =  f_\text{single-shot}(U) \UVSpiderDiagram  \label{eq:single_shot_shadow}
\end{align}
and write $\hat{M^\dagger} \Vec{\phi}_s$ as the average over shadows projected to the different basis elements
\begin{align}
    \frac{1}{N_{\rm shots} }\hat{M^\dagger} \Vec{\phi}_s = 
    \underset{U\in \mathcal{U}_s}{\mathbb{E}}\left[ \mathcal{S}_\text{single} \right] \circ \vec{{\rm \mathbf{B}}}
    \label{eq:projected_avr_shadows}
\end{align}


In the ideal case, i.e. for an infinite number of measurements and in the absence of noise, $\vec{v}_{\rm op}$ coincides with the exact decomposition of the environment tensor, $v_{\rm exact}$ and the multiplication by $\hat{M}$ gives an exact evaluation of the cost function
\begin{align}
    (\hat{M}\cdot { \vec{v}_{\rm exact}})_i &= \sum_{j} \left(\TdiagB{B_j}{U_i}{U_i} \times { {v_{\rm exact}}_j}\right)\\
    &= \TdiagB{E}{U_i}{U_i} = f_\text{exact}(U_i)
\end{align}
In contrast, for any finite number of measurements the connection between measurements and the overlap matrix is not exact: $\vec{v}_{\rm op}$ is an unbiased estimator of the amplitude vector.

An important outcome of the linear-regression formalism is an estimate of the average accuracy of the extracted environment tensor, derived in appendix \ref{app:reg}, which corresponds to the trace norm of the inverted second-moment matrix
\begin{align}
    {\rm Var}_{\rm tomo}(\vec{v}_{\hat{\rm E}}) &= {{\mathbb{E}}\left[\norm{\vec{v}_{\rm \hat{E}} - \vec{E}}^2\right]}  \label{eq:linear_bound}\\
    &= \frac{1}{N_{\rm shots}} {{\rm Tr}\left(\left(\frac{1}{N_{\rm shots}}\hat{M}^\dagger \hat{M}\right)^{-1}\right)} {\left< {\sigma_f(U)}^2\right>_U} \nonumber
\end{align}

This approximation assumes that the  gates are randomly chosen, in a way that is uncorrelated from the value and the variance of the cost function measurements. The formula includes three factors: the first factors $1/N_{\rm shots}$ is responsible for the inverse-linear scaling of the variance with the number of shots. The second factor, ${{\rm Tr}(N_{\rm shots}(\hat{M}^\dagger \hat{M})^{-1})}$, determines the dependence of the variance on the ensemble of gates, given by the trace of the inverse of the second-moment matrix. This term is the most significant in the design of gate ensemble for tomography.
The third and last factor is the variance of a single-shot measurement of the cost function, averaged over the ensemble of different gates. 

In order to minimize the variance for a fixed number of gates, assuming we have no prior knowledge of the environment tensor, the unitary gates should be chosen in a configuration that avoids small singular values of the second-moment matrix, keeping the second term small. As shown in Appendix \ref{app:reg}, this quantity is minimal for unitary 2-design. Alternative techniques are suboptimal in terms of the number of measurements but, as we will explain, may have other practical advantages.

We next analyze two algorithms implementations of tomography gate sets. The first algorithm, the uniform landscape tomography, sample gates from a uniform distribution of gates, either from a discrete unitary 2-design or from the continuous Haar-random unitary distribution. A similar approach which uses uniform shadows was used in quantum state tomography \cite{aaronson2018shadow, huang2020predicting, nguyen2022optimizing} and, more recently, in quantum channel tomography \cite{kunjummen2023shadow, levy2021classical, roth2018recovering}. The second algorithm is based on Clifford tableaux and includes a small subset of Clifford gates. This approach uses an explicit construction of gate measurements that allows for independent measurements of the environment tensor in separate subgroups.



\subsection{Uniform landscape tomography\label{sec:uniform}}

A natural implementation of our environment tomography uses a uniform distribution of random unitaries, sampled from either the continuous Haar-random unitary distribution, or from a discrete set of matrices that form a unitary 2-design. This approach, which closely resembles the algorithm of shadow tomography of quantum states, fulfills the optimal bound in terms of the number of shots for general environment tensor. Although generally inefficient in the number of unique circuit measurements, it gives additional insight into the roles of the different parts of linear tomography. In this section, we give explicit calculations for both the reconstruction formula and the variance of the final reconstructed tensor for this special case. 

By choosing uniformly distributed gates, such as unitary 2-design, we can calculate explicitly the second moment matrix $M^\dagger M$ using Haar integration (see Appendix \ref{app:opti_2_design} for details):
\begin{align}
    \frac{1}{N}(M^\dagger M)_{i,j} &= \int_{\rm Haar}{\TdiagB{B_i}{U}{U}\;\TdiagB{B_j}{U}{U} \;dU}\\
                        &= \begin{cases}
                            1 &\text{$i=j=1$}\\
                            \frac{1}{d^2 - 1} &\text{$i=j>1$} 
                            \end{cases}
    \label{eq:uniform_varmat}
\end{align}
where $d = 2^k$ is the dimension of the Hilbert space.

Likewise, we can perform the integration over the single-shadow averaging in the right-hand side of the Eq.~(\ref{eq:MtMa}). Here, the single-shot shadows contain stochastic noise that contribute to the variance of the final tensor reconstruction. 
As the number of shots increases, the level of shot noise decreases, and the averaged shadow approaches the expected value of the shadows across numerous experimental realizations. We define the classical shadow as the a shadow calculated using a the expected cost function value under the limit of many single-shot measurement:
\begin{align}
    \mathcal{S}_{\rm exact}(U) &= \E[\mathcal{S}_\text{single-shot}(U)]\\
    &= f_{\rm exact}(U,U^\dagger)\UVSpiderDiagram\\
    & = \left(\Tdiag{U}{U}\right)\UVSpiderDiagram
    \label{eq:Shadow_exact}
\end{align}

For a large number of shots we can average over the uniform distribution of the exact shadows using Haar-integration \cite{novaes2014elementary,kostenberger2021weingarten,collins2022weingarten,mele2024introduction}
\begin{align}
    \E_{\substack{ U \in U(d) \\ \text{ shots}}}[&\mathcal{S}_\text{single-shot}] = \int_{\rm Haar}{\mathcal{S}_{\rm exact}(U) dU } \\
    &= \frac{(d^2-2)}{d^2(d^2-1)}\vcenter{\hbox{\includegraphics[scale = 0.4]{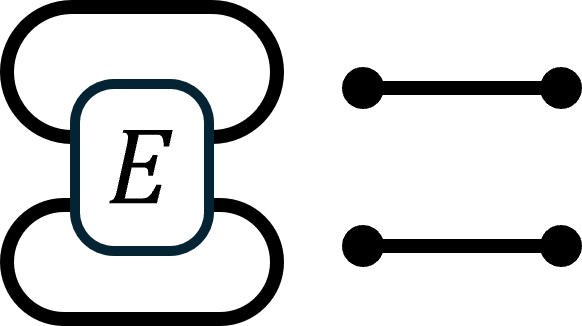}}} + \frac{1}{d^{2} - 1}\vcenter{\hbox{\includegraphics[scale = 0.4]{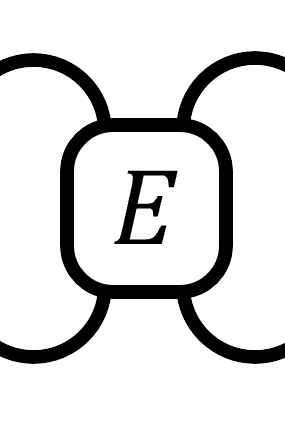}}}
    \label{eq:Shadow_tomography}    
\end{align}
In the resulting expression, the first term is an identity tensor which acts as a constant function under symmetric contraction of unitary gates. The second term is a reconstruction of the environment tensor, shrunk by a prefactor of $1/(d^2 - 1)$.
From Eq.~\ref{eq:projected_avr_shadows} we can write the resulting approximation for the vector $\hat{M^\dagger} \Vec{\phi}_s$ in the limit of noiseless measurements
\begin{align}
    \frac{1}{N_{\text shots}}(\hat{M^\dagger} \Vec{\phi}_s)_i \approx  \frac{d^2-2}{d^2-1}E_1\delta_{i1} + \frac{1}{d^2 - 1} E_i
\end{align}
where $E_i$ is the $i$-th coordinate of the environment tensor in the basis representation 
$E_i = \EBdiag$.

One can verify that in the limit of noiseless measurements we recover the original tensor using Eq.~\ref{eq:inv_solution}: Applying the inverse of variance matrix on the averaged shadows vector $\hat{M^\dagger} \Vec{\phi}$ corrects the prefactors of both the environment term and the constant background. Note that the expression in Eq.~\ref{eq:Shadow_tomography} is very similar to the quantum state shadow tomography  \cite{ huang2020predicting, nguyen2022optimizing}, an algorithm designed to predict properties of local operators using measurements of classical shadows of the density matrix. Similarly, in our algorithm we measure multiple projections of the environment tensors, store them as classical shadows, and recover the environment tensor by averaging over the shadows. 

The variance of the recovered environment tensor can calculated from Eq.~\ref{eq:linear_bound} and equals to
\begin{align}
    {\rm Var}_{\rm tomo}(\vec{v}_{\hat{\rm E}}) &= \frac{1+(d^2-1)^3}{N_{\rm shots}} {\left< {\sigma_f(U)}^2\right>_U}.\label{eq:var_tomo}
\end{align}
Because the eigenvalues of the variance in Eq.~\ref{eq:uniform_varmat} is near-uniform, unitary 2-design are optimal for environment tomography, in analogy to the case of quantum channel tomography \cite{scott2008optimizing} (See Appendix \ref{app:opti_2_design}). We conclude that Eq.~(\ref{eq:var_tomo}) is a lower bound for the variance of environment tomography.


The disadvantage of this approach is that the number of gates that need to be probed is very large, and cannot be reduced further than the minimal size of a complete unitary 2-design. In what follows, we develop an alternative method that relaxes the constraints on the set of gates and allows for more flexible gate set designs that are advantageous from a practical perspective.

\subsection{Tableaux-based tomography}
\label{sec:stab}

A complementary approach to the uniform sampling strategy is choosing a gate-set design that deliberately measures and extracts specific components of the environment tensor. As we explain below, this strategy is convenient in the presence of physical limitations that limit the number or precision of the probed circuits. Here we suggest a constructive approach that relies on the Clifford tableaux formalism, leveraging the properties of Clifford gate conjugation on Pauli strings. This enables us to isolate small subsets of amplitudes through a series of circuit measurements.


To resolve different components of the environment tensor we represent it using the horizontal Pauli basis (see Appendix \ref{app:decomp})
\begin{align}
    \Tderivderiv{E} = \sum_{i,j} e_{i,j} \ABUVdiagClean{P_i}{P_j}
    \label{eq:eij}
\end{align}
where $P_i$ and $P_j$ are $k$-qubits Pauli strings and $i,j \in 1,2,... 4^k$. 
In this basis the cost function evaluated on $U$ is written as the sum of traces
\begin{align}
    \Tdiag{U}{U} &= \sum_{i,j} e_{i,j} \ABUVdiag{P_i}{P_j}{U}{U} \\
    &= \sum_{i,j} e_{i,j}\Tr(P_i U P_j U^\dagger)
    \label{eq:eij_u_udag}
\end{align}
Each trace in the expression above can be viewed as an overlap of the conjugated Pauli string $U^{\dagger} P_i U$ with the Pauli string $P_j$.
Combining all terms, the action of the environment tensor on a given unitary gate can be reinterpreted as calculating the transformation of all Pauli strings under conjugation with the gate, and summing the weights of the different outcomes given by the environment tensor amplitudes.
This calls naturally for examining the evaluation of the cost function on Clifford gate.


Clifford gates are defined as normalizers of the Pauli group, i.e. operators that map Pauli strings to Pauli strings  under conjugation, up to a complex phase. 
Each Clifford gate can be represented using a Clifford tableau, describing how different Pauli strings are transformed under gate conjugation \cite{gottesman1998heisenberg, aaronson2004improved}. 
To describe the complete transformation of Pauli strings induced by a Clifford gate conjugation, it is sufficient to describe the transformation of selected $2k$ generators which span all Pauli strings, conventionally chosen as the X and Z gates on each qubit. The tableau then describe the Pauli strings to which each generator is mapped.

For example, the CNOT gate is a Clifford gate that transforms 2-qubit Pauli strings as follows \footnote{For the purposes of this paper, we will not use the table notation used in ref. \cite{aaronson2004improved} and will list the transformation outcome as a list.}: 
\begin{align}
    {\rm CNOT} = 
    \begin{Bmatrix} 
        Z_1& \rightarrow &Z_1 I_2 \\
        X_1& \rightarrow &X_1 X_2 \\ 
        Z_2& \rightarrow &Z_1 Z_2 \\
        X_2& \rightarrow &I_1 X_2.
    \end{Bmatrix}
    \label{eq:CNOT_transformation}
\end{align}
Using these rules, one can derive the transformation of any arbitrary Pauli string by chaining the different transformation of each component of the Pauli string.

When a Clifford gate $U_0$ is applied to Eq.~(\ref{eq:eij_u_udag}) it leads to a linear superposition of all $4^k$ amplitudes $e_{i,j}$,
with $(i,j)$ satisfying $P_i = U_0 P_j U^{\dagger}_0$. To extract the individual elements $e_{i,j}$ we consider the set of $4^k$ gates of the form $U_m = P_m U_0$ where $P_m$ is one of the $4^k$ Pauli strings.
%
%
All the gates in this set probe the same $4^k$ components $e_{i,j}$ with equal magnitude, but with a unique pattern of signs. When arranged in a matrix, these amplitudes are equivalent to a basis transformation from the basis of projections on the set of unitary gates to the horizontal Pauli basis, which is equivalent to $2^{2k}H^{\otimes 2k}$, up to permutation of rows and columns:
\begin{align}
\begin{pmatrix}
  f(P_1 U_0) \\
  f(P_2 U_0) \\
  \vdots \\
  f(P_{4^k} U_0)
\end{pmatrix} = 2^{2k} H^{\otimes 2k}
    \begin{pmatrix}
  e_{i_1,j_1} \\
  e_{i_2,j_2} \\
  \vdots \\
  e_{i_{4^k},j_{4^k}}
\end{pmatrix}.
\end{align}
The equation above is yet another special case of Eq. \ref{eq:linear_reg} for a subset of environment elements, where the left hand size is the measured samples vector $\phi_s$, and the Hadamard transformation is the overlap matrix $\hat{M}$. More fundamentally, conjugation under the set of gates described above is  related to the subset of the horizontal Pauli basis up to a linear Hadamard transformation 
\begin{align}
\begin{pmatrix}
  \;\;\UVSpiderDiagramWide{P_1 U_0}{U_0^\dagger P_1^\dagger} \;\;\;\; \vspace{3mm}\\
  \;\; \UVSpiderDiagramWide{P_2 U_0}{U_0^\dagger P_2^\dagger} \;\;\;\; \\
  \vdots \\
  \;\; \UVSpiderDiagramWide{P_{4^k} U_0}{U_0^\dagger P_{4^k}^\dagger} \;\;\;\;
\end{pmatrix} = 2^{2k} H^{\otimes 2k}
    \begin{pmatrix}
  \ABUVdiagClean{P_{i_1}}{P_{j_1}} \vspace{3mm}\\
  \ABUVdiagClean{P_{i_2}}{P_{j_2}} \\
  \vdots \\
  \ABUVdiagClean{P_{i_{4^k}}}{P_{j_{4^k}}}
\end{pmatrix}.
\end{align}

By measuring the whole set we can invert the transformation matrix and extract the amplitude of each element individually.
Once we are able to measure and recover batches of $4^{k}$ amplitudes, we need to probe different sets of gates, i.e. different Clifford gates $U_0$, until we recover all the measurable elements of the environment tensor. 
Due to the nature of the Clifford group, the different sets inevitably have non-empty overlap, where the primary example is the identity element, $P_i=P_j=I$, which appears in all sets. This overlap along with other repeating Pauli pairs introduce inefficiencies in measurements of the tensor elements, as some elements are measured more frequently than others. 
In principle, repetitions could be avoided by relying on previously extracted information to reduce the number of circuit evaluations, however, the final accuracy of the reconstruction will be reduced due to inaccuracies carried from one subset to the evaluation of the others. 
Instead, we use the overcomplete set of measurements to increase the accuracy of the overlapping elements. Each group of measurements is performed independently, and values of the matrix elements that have been computed more than once are averaged over the different realizations.

We now search for a minimal cover of gate sets that probes all the components of the environment with minimal overlap. This task can be performed algorithmically, by going over different configurations of sets and searching for an optimal cover.
For 1-qubit gates, the optimal configuration includes 3 batches of 4 gates each, for a total of 12 gates. As expected, this number is larger than the total number of $e_{i,j}$ amplitudes that equals $N_{\rm relevant}=10$, see Eq. \ref{eq:relevant_dof} with for $k=1$. For 2-qubit gates, the minimal number of sets we found is 17, which adds up to 272 different circuits, about 20\% more than the number of $e_{i,j}$ amplitudes for $k=2$, which equals 226. Importantly, unlike uniform shadow tomography with Haar-random gates, the number of circuits used by the tableaux-based method does not grow with the number of shots.  


The combined subsets of gate substitution also fit nicely into the formalism of linear inversion tomography. In this case, the matrix of overlaps is a combined sum of Hadamard matrices embedded sparsely into small selections of rows and columns - according to the set of probed Pauli components. Given the overlap matrix, the performance of the algorithm can be theoretically estimated as well from Eq. 
\ref{eq:linear_bound} (see Appendix \ref{app:stab} for more details).
This is a special case of regression tomography, in which the sampling gates are not evenly distributed, and has the advantage of using a minimal, or close to minimal, number of unique circuits.

\begin{figure}
    \centering
    \includegraphics[width=\columnwidth]{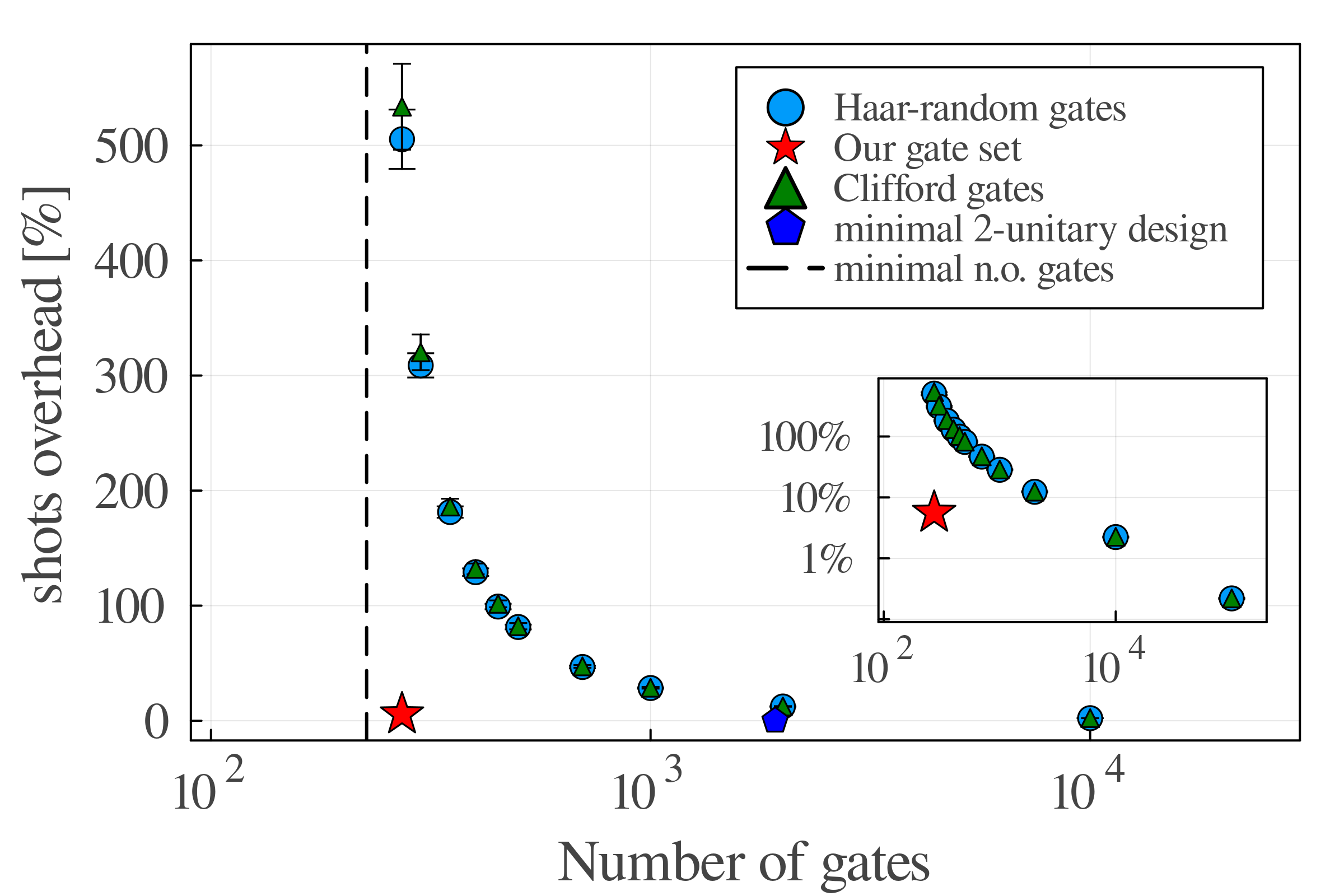}
    \caption{Expected overhead in number of shots by using different gate sets as a function of the number of gates. We compare our gate set with sets of 2-qubit Haar-random unitary gates, sets of random Clifford gates, and the smallest known unitary 2-design set.}
    \label{fig:gate_set_benchmark}
\end{figure}

To estimate efficiency of our tableaux-based approach, we compute the overhead introduced by this method, with respect to a random choice of Clifford gates or Haar-random gates. The overhead is quantified by the prefactor of the linear regression estimation in Eq. \ref{eq:linear_bound}, evaluated for single shot per gate-
\begin{align}
    {\rm Var}_{\rm single} = {{\rm Tr}\left[N_{\rm gates} \left(\hat{M}^\dagger \hat{M}\right)^{-1}\right]}
\end{align}
The result of our analysis are shown in Fig.~\ref{fig:gate_set_benchmark} 
and demonstrate that our method can reach an overhead of 5.8\% using only 272 gates, whereas a set of uniformly sampled gates from either the Haar-random distribution or the Clifford set arrive the same overhead with about 10 times the number of circuits. Conversely, the minimal set of unitary 2-design for 2-qubit gates of Ref.~\cite{gross2007evenly}, while demonstrating no shot overhead, requires about 7 times more circuit than our designed set of gates.



\subsection{Tomography in presence of hardware constraints \label{sec:constr}}
A key advantage of tableaux-based tomography is that this approach offers full control over the type of gates used for full tomography. In particular, we can find an upper bound for the number of elementary gates required for full tomography. Minimizing the number of entangling gates for the tomography process is important for performing calculations on real quantum hardware where each multi-qubit entangling gate generates non-negligible error, and using less of them allows for a more accurate implementation of the tomography algorithm. In the following analysis we focus on cases where the native gates include single-qubit gates and the CNOT gate as a native entangling gate, as used for example in IBM quantum computers.

Here, we considered two distinct cases: the case of all-to-all connectivity model where each CNOT gate can be applied between any pair of qubits, and the case of linear connectivity case where CNOT gates are applied only between nearest neighbors of a linearly aligned array of qubits. 
We then track the way CNOT gates act in conjugation on Pauli strings, and how they can be used to transform one Pauli string to another; see Appendix \ref{app:CNOTCount} for details. We find that the minimal number of sequential CNOT gates required to probe all degrees of freedom of a $k$-qubit gate is \begin{align}
    N^{\rm min}_{\rm CNOT} = \begin{cases}
        k ,& \text{all-to-all connectivity}\\
        2k-2 ,& \text{linear connectivity}
    \end{cases} \label{eq:CNOT_count}
\end{align}
In both cases the number of CNOTs scales linearly with $k$, and is therefore much smaller than the number of CNOTs required to implement a general $k$-qubit gate, which is bounded from below by  
$N^{\rm general}_{\rm CNOT} = (4^k - 3 k - 1)/4 \gg N_{\rm CNOT}^{\rm min}$ \cite{shende2004minimal}.
For $k=2$ only two CNOT gates are sufficient to perform a full tomography using our tableaux-based approach, while three gates are needed to implement a generic 2-qubit gate. This finding is corroborated by the explicit list of gates provided in appendix \ref{app:stab}, which can be implemented using two CNOT gates only.

A complementary question that can be addressed using our approach is the identification of the amplitudes that can be probed using {a smaller number of CNOT gates, $N_{\rm CNOT}<N^{\rm min}_{\rm CNOT}$}. 
For example, by applying gates that correspond to the tensor product of 1-qubit gates, one cannot probe Pauli strings that pair a non-identity Pauli matrix to an identity Pauli matrix for any of the qubits. In this case, the total number of measurable components is then 
\begin{align}
    N_{\rm measurable}^{{N_{\rm CNOT} = 0}} = 10^k \label{eq:no_meas0}
\end{align}
More generally, for $k$-qubit gates limited to $t$ CNOT gates (with all-to-all connectivity), the number of measurable degrees of freedom is given by:
\begin{align}
    N_{\rm measurable}^{{N_{\rm CNOT}} = t} =  2 + \sum_{l=0}^{t}{6^l {\binom{k}{l}} (10^{k-l}-0.5^{l-1})}.\label{eq:no_meas}
\end{align}
As expected, for $t=N_{\rm CNOT}^{\rm tableaux}$ Eq.~{\ref{eq:no_meas}} returns the total number of measurable amplitude, $N_{\rm relevant}$, given by Eq.~\ref{eq:relevant_dof}. For 2-qubit gate, $k=2$, we obtain $N_{\rm measurable}^{{N_{\rm CNOT} = 0}} = 100$, $N_{\rm measurable}^{{N_{\rm CNOT} = 1}} = 208$ and $N_{\rm measurable}^{{N_{\rm CNOT} = 2}} = 226$.
For each value of $n$, it is possible to construct a cover of sets that probes only the measurable components using the algorithm described above. See more details in Appendix \ref{app:CNOTRestrict}.
These general relations are useful for possible extensions of the optimization algorithm in the spirit of the adaptive ansatz strategy\cite{grimsley2019adaptive,tang2021qubit}. For example, by performing gradient tomography on large blocks of 3- or 4- qubit gates while using a predefined CNOT gates budget, one can explore different structures of entangling gates while limiting the number of circuit evaluations.

\section{Finding the minimum of the environment tensor}
\label{sec:minimum}

The landscape tomography described in the previous section allows one to express the quantum problem of gate optimization as a classical bilinear cost function. By solving the optimization problem, one can then find the optimal gate that minimizes the value of the cost function.
Finding analytical expression for the optimal gate is a difficult problem, and so we instead resort to numerical optimization algorithms such as gradient descent on unitary manifolds \cite{abrudan2008steepest, hauru2021riemannian, wiersema2023optimizing} or "gate-by-gate" optimization subroutine in which we use consecutive linearization of the cost function to optimize each gate individually, iterating from $U$ to $U^\dagger$ \cite{haghshenas2021optimization}.
After the optimal unitary $U_{\rm opt}$ has been found, the old gate is replaced with the optimized one and the algorithm move to the next gate- characterizing its environment tensor and finding the optimal gate. 
The optimization algorithm is finished when the cost function value no longer improves, or when the maximal number of shots is reached.

For optimization using the steepest descent algorithm, the first and second derivatives can be calculated directly across the whole landscape using the reconstructed environment tensor:
\begin{align}
    &\frac{\partial f_{\rm exact}(U)}{\partial U} = \Tderiv{U}\\
    &\frac{\partial^2 f_{\rm exact}(U)}{\partial U \partial U^\dagger} = \Tderivderiv{E}
\end{align}
Using landscape tomography for single-gate optimization has an advantage over using gradient descent with measurements of local derivatives. The former approach gives a full description of the cost function dependency on a single gate. This global description of the landscape can be used to calculate the local derivatives for any gate substitution without carrying any further measurements, unlike gradient-based optimization methods in which measurements of local gradients cannot be reused.  
Additionally, local measurements of second-order derivatives do not retrieve full global information about the single-gate cost function landscape, as they naturally measure only the part of the environment tensor that is projected onto the local tangent space of the unitary manifold. As a result, measurements of the Hessian of the cost function at one point cannot be used for later steps of the optimization method.

\section{Numerical results \label{sec:numerical}}

To test the performance of environment tomography algorithms, we performed numerical simulations of landscape tomography on parametric quantum circuits. As a concrete example, we implemented the VQE algorithm on the Ising-model Hamiltonian $H = \protect\sum_{n}{J_z Z_n Z_{n+1} - h_x X_n}$, with a magnetic field of $h_x = 0.5$ and open boundary conditions. For our variational state we chose a dense-encoding ansatz with 3 staircase layers, where each layer is constructed of generic 2-qubit gates. The cost function is evaluated using two Pauli string measurements, from which the expectation values of all the Hamiltonian terms can be extracted- $X^{\otimes n}$ for the transverse field and $Z^{\otimes n}$ for measuring the $ZZ$ couplings. 
The numerical simulations were performed using tensor-networks, allowing for simulations of circuit sampling with finite number of shots as well as exact calculation of expectation values and environment tensors using tensor contractions. 

\begin{figure}
    \centering
    \includegraphics[width=\columnwidth]{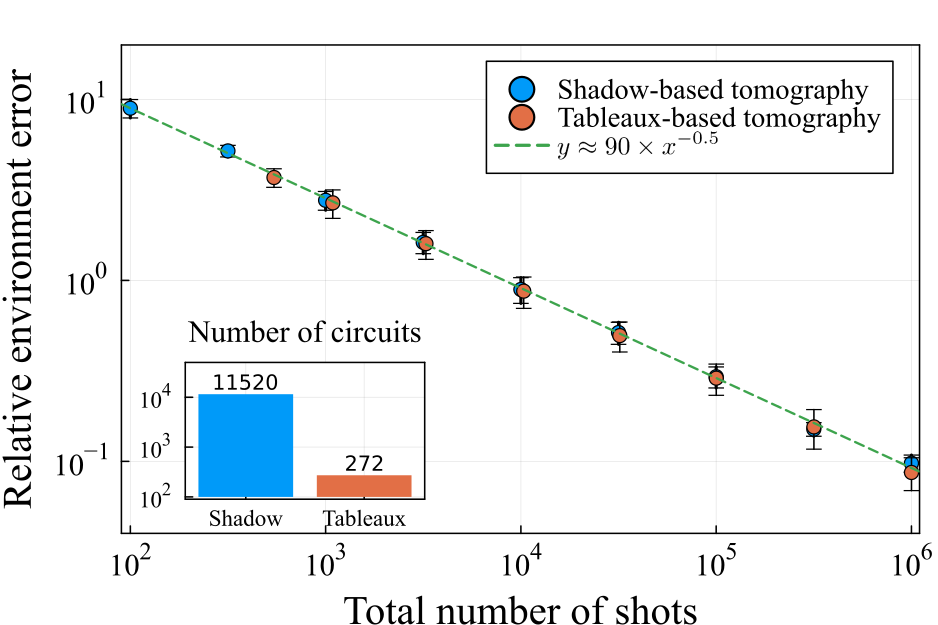}
    \caption{ The accuracy of the reconstructed environment (Eq. \ref{eq:error_env}) as a function of the number of shots plotted for two tomography methods: uniform tomography, and tableaux-based tomography. The calculation were performed for Ising model VQE circuit, with random ansatz initialization of 2 staircase layers.}
    \label{fig:env_fidel}
\end{figure}

\begin{figure}
    \centering
    \includegraphics[width=\columnwidth]{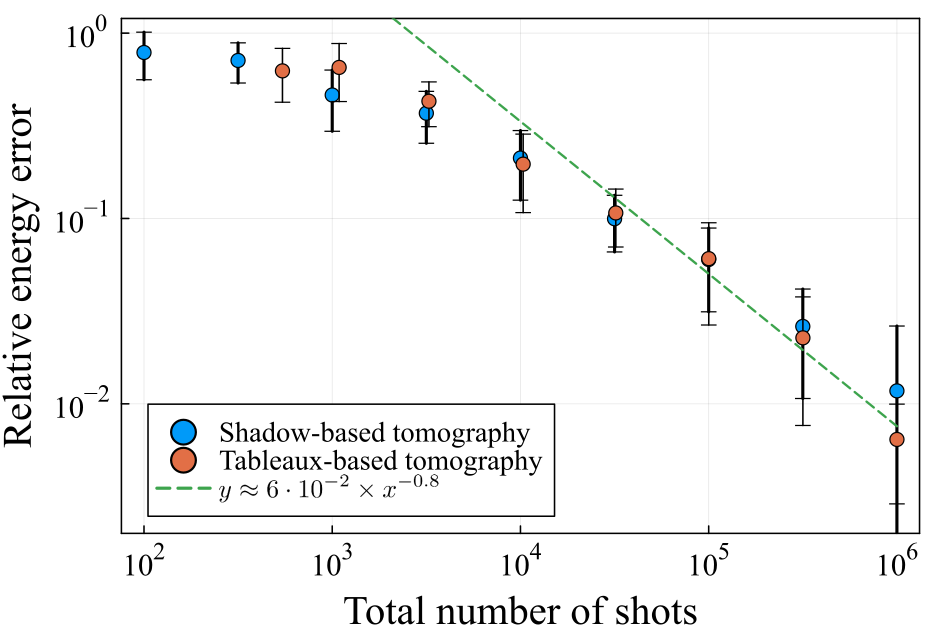}
    \caption{ The relative energy error of the optimal gate as a function of the number of shots.}
    \label{fig:U_fidel}
\end{figure}

\begin{figure}
    \centering
    \includegraphics[width=\columnwidth]{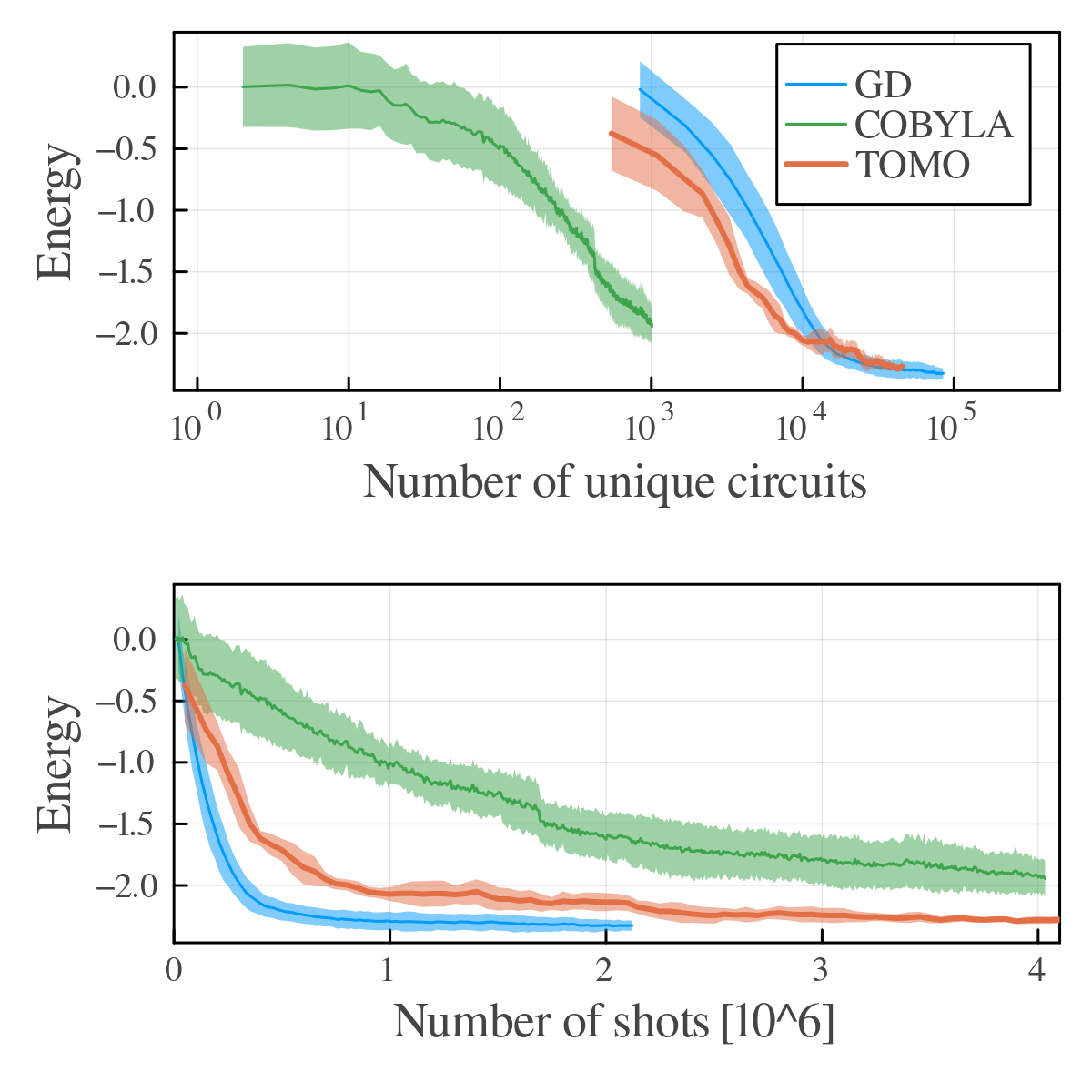}
    \caption{ A demonstration of our optimization algorithm for VQE of the Ising Hamiltonian. The energy of the prepared state is plotted against the number of shots performed during the optimization algorithm.}
    \label{fig:opt_res}
\end{figure}


As a first test of our algorithm, we characterized the environment tensor of a single gate while keeping the other gates fixed to random values using landscape tomography, and compared it to the exact environment of the gate. In Fig.~\ref{fig:env_fidel}, we plot the accuracy of the reconstructed environment tensor as a function of the number of shots {for 6 qubits ansatz with 3 stair case layers,}  for the two methods studied in this paper: uniform landscape tomography and tableaux-based tomography. The environment is The shadow- based tomography methods used 11520 distinct unitary circuits to sample all  Clifford 2-qubit gates, while tableaux-based tomography used a smaller set of 272 gates (17 groups of 16 gates) as described in Section~\ref{sec:stab} and Appendix \ref{app:stab}, and included only gates that contained up to 2 CNOT gates. 
The accuracy of the reconstruction was quantified by the average distance between the original and the reconstructed reduced cost function for random unitary substitutions, normalized by the mean value of the cost function, as follows
\begin{align}
    {\delta ({\rm energy}_{\rm avg})}(E) = \frac{\norm{\Tdiag{U}{U} - f_{\rm exact}(U)}_{U\in \mathcal{U}_{\rm check}} }{\norm{f_{\rm exact}}_{U\in \mathcal{U}_{\rm check}}} \label{eq:error_env}
\end{align}
We found that the two methods have similar performance, and the differences are statistically insignificant. 
The curves follow the scaling $\thicksim N_{\rm shots}^{-0.5}$, as expected from shot noise errors. Despite of the similarities, the tableaux-based tomography methods offers a practical advantage over the others because it requires fewer unique circuits and having a reduced CNOT gate count, minimizing decoherence errors caused by multiqubit gates.

Having characterized the environment tensor, we searched for the optimal gate that minimizes the reconstructed reduced cost function using a gradient descent optimization algorithm. Naturally one expects an accurate reconstruction of the environment tensor to yield a better estimation of the optimal unitary gate, resulting in cleaner optimization steps. It is therefore important to check how efforts to extract an accurate environment tensor translate into the accuracy of the optimized minimal unitary. We quantify the relative error in the value of the cost function (the Hamiltonian energy in our case) between the correct optimal unitary and the ones calculated from the reconstructed environment according to 
\begin{align}
    {\delta ({\rm energy}_{\rm opt})}(E) = \frac{\abs{\TdiagWide{U_{\rm opt}}{U_{\rm opt}} - {\rm min}(f_{\rm exact})} }{{\rm min}(f_{\rm exact})} \label{eq:error_ener}
\end{align}
In Fig.~\ref{fig:U_fidel}, we plot the relative error in the energy as a function of the number of shots, 
along with a numerical fit of a power-law scaling. We observe that although the exact value of the error is noisy, the error in energy scales numerically according to the power law ${\delta ({\rm energy}_{\rm opt})} \thicksim N_{\rm shots}^{-0.8}$. Because the error is quadratic in the gate operators, we expect that at larger numbers of shots, the error should be proportional to  the inverse of the number of shots, $\thicksim N_{\rm shots}^{-1}$. Our numerical result is also an indication that a smaller number of shots may be sufficient for an optimization step then for accurate full-environment reconstruction, as the optimal gate of the reconstructed environment tensor is less susceptible to noise than the full environment itself for large number of shots. 


Finally, we studied the performance of the optimization algorithm for finding the ground state of the Ising model of 8 qubits with a full run of the VQE algorithm with an ansatz of 2 staircase layers. We compared 3 optimization methods: steepest descent -- a gradient-based optimization based on the parameter-shift rule (GD) \cite{sweke2020stochastic}, gradient-free optimization based on the COBYLA optimization algorithm \cite{powell2007view,lavrijsen2020classical}, and our method of consecutive landscape tomography and gate replacements (TOMO). These three methods present different approaches for utilizing the limited amount of shots available to extract as much information as possible about the cost function and optimize it rapidly and efficiently. 
For the gradient descent algorithm, we parameterized each 2-qubit gate with 15 parameters \cite{shende2004minimal} and performed the optimization for the different rotation angles. At each iteration we calculated the full gradient of all parameters using the parameter-shift rule and step in the direction of the gradient using a learning rate of $\eta = 0.15$. We allocated 200 shots per parameter (100 for each parameter-shift rule circuit evaluation) for each gradient evaluation using parameter-shift rule.
For the COBYLA optimization method we allocated 10,000 shots per cost function evaluation, while for our method we allocated 50,000 shots for the full environment tomography of each 2-qubit gate.  The results were averaged over 8 realization of the algorithm. Fig. \ref{fig:opt_res} compares the performance of different optimization algorithms by plotting the energy of the optimized state as a function of number of unique circuits in Fig. \ref{fig:opt_res}(a) and as a function of number of shots spent for optimization in Fig. \ref{fig:opt_res}(b).  



Gradient-based (GD) and gradient-free (COBYLA) optimization illustrate two different resource-allocation regimes. Gradient descent can converge with a relatively low number of shots, but each update requires evaluating many parameter-shift circuits, leading to a large number of unique circuit evaluations. In contrast, COBYLA relies on fewer circuit structures per iteration, but typically requires many more shots to reach convergence. Our algorithm occupies an intermediate regime: it leverages the reconstructed local landscape to extract more information from each gate update, achieving faster convergence than COBYLA in terms of shots while requiring fewer circuit evaluations than naive gradient-based updates per iteration. These results highlight the trade-offs between different optimization strategies in a realistic circuit-optimization setting.







\section{Comparison with parameter-shift rule\label{sec:compare}}

Having introduced the essentials of our method, we now discuss its relation to the common parameter-shift rule approach. A natural alternative to environment tomography consists of applying a second-order parameter-shift rule to measure the full Hessian of the reduced cost function, using an explicit full parameterization of the unitary gate. Since the cost function is second order in $U$, and the second derivative in $U$ is constant throughout the entire function landscape, it is tempting to conclude that the measured Hessian is equivalent to the measured environment tensor.
In reality, this is not the case, as a single local full Hessian does not contain enough information about the environment tensor by its own. This difference is quite easy to spot when counting the number of degrees of freedom in the Hessian, which is a symmetric matrix, against the number of the environment components. 
It is therefore essential to measure the Hessian around at least two different gate substitutions.
An additional comparison can be made for the minimal number of circuits required to measure the Hessian and the environment tensor. A naive characterization of the Hessian requires measuring nearly $4\times(4^k-1)^2$ different circuits for full unitary $k$-qubit gate with $4^k-1$ parameters, while environment tomography requires only $\thicksim (4^k - 1)^2 + 1$ circuits. We conclude that our algorithm offers a new way to evaluate the Hessian of a full $k$-qubit gate in parametric quantum circuits using a reduced number of quantum circuits.
The tableaux-based tomography has the additional advantage of the ability to control the type of gates to use only limited number of CNOT gates for example, as portrayed in Eq. \ref{eq:CNOT_count}.

In fact, our method is closely related to the general parameter-shift rule method \cite{wierichs2022general, izmaylov2021analytic, kyriienko2021generalized}, which is used to calculate the gradient of a multi-qubit parameterized gate $U(\theta)$ with a wide spectrum of eigenvalues or rotational frequencies. According to the general parameter-shift rule, a set of measurements with different $\theta$ values are performed to reconstruct the values and gradients of the multi-qubit gate, in a way similar to discrete Fourier transform. Our method of landscape tomography and the general parameter-shift rule method share many common features - both obtain a full description of the cost function landscape from a set of measurements and both method yield optimal accuracy for equidistant choice of thetas or unitary gates (which is the unitary 2-design). Both methods are also used for parameter-wise optimization (or the Rotosolve algorithm). The general parameter-shift rule is suited for problem-inspired ansatzes like QAOA, which typically associate one parameter with a large many-qubit gate block. Our work, on the other hand, fits best for cases where hardware-efficient or adaptive ansatz are used, and works with a full gates that are densely parameterized, taking advantage of the structure of the full unitary gate to save resources for multi-parameter derivative characterization.

\section{Conclusion and future directions \label{sec:conclude}}
In this paper we presented a new method for the optimization of densely parameterized quantum circuits using a gate-by-gate optimization technique: each gate is optimized by first determining the full dependence of the cost function on the gate, a process referred to as quantum landscape tomography, and then using classical optimization techniques to find the optimal gate.
We developed a general framework for landscape tomography that allowed us to theoretically analyze the efficiency of different sets of gates. We specifically focus on
random gates (sampled from either the Haar random unitary distribution or a unitary 2-design set), as well as a specially designed set of gates obtained by a tableaux-based approach. The freedom of choice of the gate set used for tomography allowed us to control and minimize the number of circuit evaluations used for landscape tomography while retaining near-optimal efficiency and accuracy of the landscape reconstruction, as well as controlling the type of gates according to restrictions imposed by the hardware such as a limited number of CNOT gates.
We demonstrated that for fully parameterized circuits, our technique offers a favorable trade-off between the number of unique circuit measurements and the total shot budget: compared with gradient descent it can reduce the overhead associated with repeated local-gradient evaluations, while compared with gradient-free optimization it can achieve convergence with fewer total shots.

The advantage of our method for the optimization of general $k$-qubit gates should be examined taking into account 
the utility of such ansatz for different quantum algorithms.  The fully parameterized gate ansatz is on the one hand natural to implement on quantum hardware given the limited connectivity most quantum devices have, the relatively low overhead of additional 1-qubit rotations, and the efficient algorithm presented in this work for full-gate optimization. On the other hand, the abundance of tunable parameters makes the ansatz harder to optimize in many cases, giving preference to deeper circuits where the parameters are sparsely distributed. This can motivate exploration of ansatz structures that contain fixed gates and layers interleaved with fully controllable $k$-qubit gates, allowing for concentrated optimization of multiple rotation angles at once while diluting the parameter count per layer. 
Another direction worth exploring is to expand the toolset available for partial tomography when strict hardware limitation applies. In section \ref{sec:constr} we described a method to cut down the number of measurements when the pool of available gates is limited to 1-qubit gates only, or any finite number of CNOT gates. A possible extension of these results involves the limitation of the search space to gates which preserve a certain symmetry \cite{larocca2022group, sauvage2022building}.

Our tableaux-based tomography shares some similarities with a recent analysis of the simulability of barren-plateau-free quantum circuits by Cerezo {\it et al.}. \cite{cerezo2023does}. This paper conjectures a connection between the absence of barren plateaus in certain VQA algorithms and the ability to classically simulate them in polynomial time. As a key example, they analyze shallow circuits in terms of the pairing between Pauli strings by conjugation, noting that a single Pauli gate can only pair to Pauli strings that are supported inside a light-cone bounded by the circuit depth. The idea of sorting the possible Pauli string pairing for a given circuit structure leads to a reduction of the variational circuit into a set of classical measurements, after which the optimization can be carried out classically. In our work we used the same logic for small sections of the unitary circuits, assuming that the action of the whole circuit might not be easily tractable. We characterized the induced pairing of Pauli strings to obtain a classical image of the cost function landscape, which we used for finding the optimal unitary classically. In essence these two works are two sides of the same coin. The same mechanism can be used for optimization of the whole circuit or only a subpart of it: one can either characterize its global behavior, as done in Ref.~\cite{cerezo2023does}, or consider the behavior of small subsection, as done here. From this perspective, our work connects the local tractability of unitary circuits to the global behavior analysis of quantum circuits described by Cerezo {\it et al.}

\section{DATA AVAILABILITY}
All data and figures that support the findings of this paper are available on request from the corresponding author. Please refer to Matan Ben-Dov at matan.ben-dov@biu.ac.il.

\begin{acknowledgments}
We would like to thank Ingo Roth, Mor Roses and David Shnaiderov for the many discussions and helpful suggestions throughout our project.
This research is supported by the Israel Science Foundation, grants number 151/19 and 154/19, by the National Research Foundation, Singapore and A*STAR under its CQT Bridging Grant.
\end{acknowledgments}


\bibliographystyle{naturemag}
\bibliography{real_bib}


\newpage

\renewcommand{\theequation}{A\arabic{equation}} 
\setcounter{equation}{0}

\renewcommand{\thefigure}{A\arabic{figure}}   
\setcounter{figure}{0}
\renewcommand{\thetable}{A\arabic{table}}   
\setcounter{table}{0}

\setcounter{page}{1}     
\pagenumbering{arabic}   

\onecolumngrid
\usetikzlibrary{decorations.markings}
\usetikzlibrary{arrows, patterns}

\appendix
\section{Properties of reduced cost function and environment tensor}
\label{app:environment_tensor}
In this chapter we give more details about the basic properties of the environment tensor. The environment tensor $E$ has a rank of $4k$ - it has two input and two output indices each of them has a dimension of $2^k$. As a results, counting all possible combinations of index values, the tensor has $2^{4k}$ complex elements which are mutually dependent. By reshaping the tensor into different shapes and dimensions we can explore several interpretations of the tensor. 
For example, the environment tensor can be either interpreted as an "inner product" function on the group of unitary matrices, or as a superoperator, a linear map on the matrix space, acting on one copy of $U$, whose result is projected onto a second copy of U:
\begin{align}
    f(U,U) = \bbra{U}\mathcal{E}\kket{U} = (U,\mathcal{E}(U)).
\end{align}

Given the physical origins of the environment tensor as an expectation value of an hermitian operator, the environment $E$ has a reflection symmetry due to the time-reversal symmetry of the orginial tensor network diagram. This reflection symmetry can be expressed through symmetry of unitary gates contraction (see figure \ref{fig:EnvTens})- $f(U,V^\dagger) = f(V,U^\dagger)^*$
\begin{align}
    \Tdiag{U}{V} = \left(\Tdiag{V}{U}\right)^*
\end{align}

In the pictorial representation of our tensor diagrams of $E$, it is illustrated as a right-left symmetry.
In the superoperator interpretation, the same symmetry is expressed as the hermicity of the superoperator $\mathcal{E}^\dagger = \mathcal{E}$

\begin{figure}
    \centering
    \includegraphics[width=.6\linewidth]{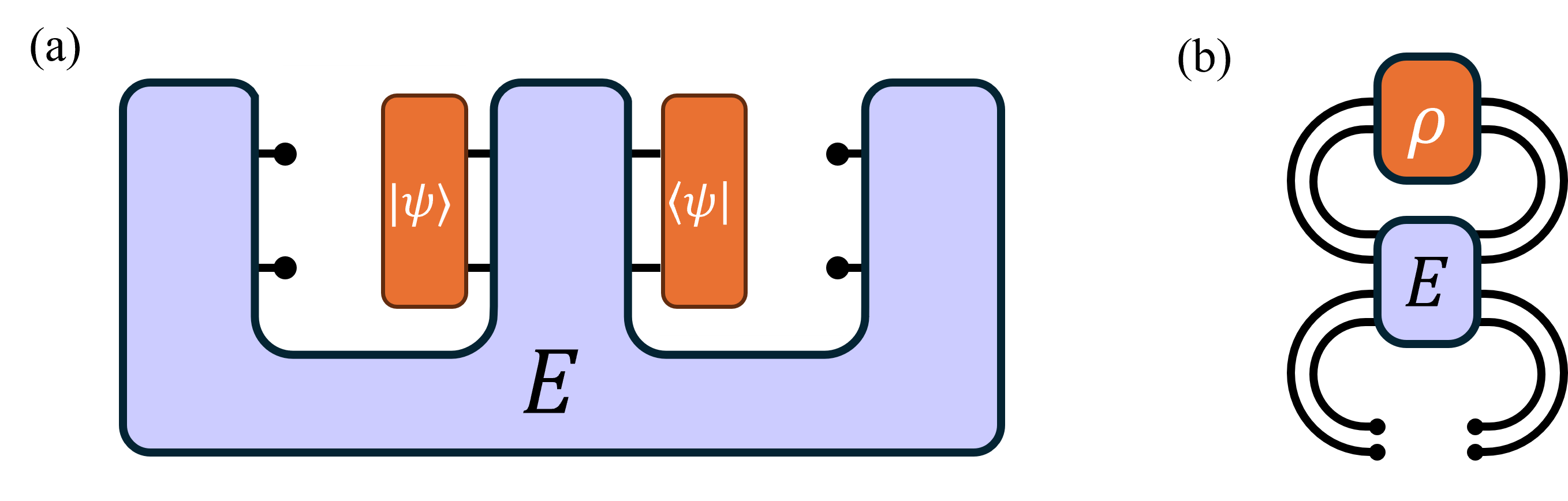}
    \caption{A re-interpretation of the environment tensor as a non-positive quantum channel. The diagrams represent (a) contracting a pure input state to the environment tensor as in mid-circuit state injection. (b) contracting a general density matrix on the environment tensor from above, contrasting the usual gate contraction to the right and left. Compare with Fig. \ref{fig:EnvTens} of the main text.}
    \label{fig:env_tens_channel}
\end{figure}

The superoperator interpretation of the environment tensor may look strange at first, as in our case it acts on unitary matrices instead of Hermitian ones, unlike in typical quantum channels. We can establish, however, a connection between these interpretations by reinterpreting the cost function in a relatively straightforward, albeit slightly contrived, manner. By treating the measured bit strings at the end of the cost function as a post-selection of the final state, we can consider the state transformation induced by contracting it to the forward indices and obtaining the output state from the backward-facing indices as a non-normalized channel with post-selection (see Fig. \ref{fig:env_tens_channel} (a)). This is reminiscent of mid-circuit injection of a quantum state drawn out of an input density matrix, while the output state is retained by non-measured ancilla-like qubits, represented by the open indices left out. The final superoperator is a weighted sum of all the possible bit strings projections, each multiplied by its corresponding energy.

This new perspective on the environment tensor takes form when treating it as a transformation function from top to bottom. Here, the input density matrix is contracted to the top indices, while the bottom indices represent the output state in the form of a density matrix, as illustrated in Fig. \ref{fig:env_tens_channel} (b). 

The resulting map is not necessarily completely-positive or trace-preserving in its current form. This limitation prevents us from assigning a direct physical interpretation of the quantum channel derived from the cost function and the quantum circuits from which the environment tensor originates. Nevertheless, it enables us to draw some similarities between the environment tensor and quantum channels.
We observe that the non-measurable elements of the environment tensor are the ones that, under the channel interpretation of the tensor, causes the density matrix mapping to misbehave. Eliminating these elements rectifies the trace-preserving property of the channel.
Although this interpretation does not satisfy all necessary conditions for quantum channels, it can still be treated as a non-convex sum of quantum channels after a few corrections, namely, rescaling of the cost function and elimination of the non-trace-preserving components of the environment tensor. The task of environment tensor tomography can then be translated into the language of quantum channel tomography, borrowing common properties and techniques such as shadow tomography and optimal projections by 2-unitary design \cite{scott2008optimizing, gross2007evenly}. 


\section{Probing the environment as linear Frame \label{app:frame}}

We showed that the using a set of unitary substitutions it is possible to retrieve the environment tensor even in presence of shot noise using linear regression. The concept of representing a state by projections on a set of overlapping vectors has been studied in linear algebra, and is represented by the concept of {\it linear frame} \cite{gross2007evenly}. Here we give a brief overview of the concept and definitions of linear frames, accompanied with their corresponding implementations for environment tensor tomography.

Linear frame is defined as a set of vectors $\{{\mathbf{e}}_k\}$ spanning the linear product space $V$, that satisfy the frame condition for every $v \in V$:
\begin{align}
    A\norm{v}^2 \leq \sum_k{\norm{({\mathbf{e}}_k, v)}^2}\leq B\norm{v}^2
\end{align}
for some finite, non-zero values $A$ and $B$. In other words, the frame is a set of vectors that cover the whole linear space and have a bounded non-zero overlap with any vector. In a way, this is a generalization of the concept of linear basis for non-orthogonal base vectors.
In our case, substitution of different unitary gates can be viewed as the projection of the environment tensor on a unitary channel, which as a whole construct a linear frame:
\begin{align}
    (\pi_U,E) = \frac{1}{d^2}\UVSpiderDiagram \circ \vec{{\rm \mathbf{E}}}
\end{align}
Given a large enough set of independent unitary operators, this frame spans the measurable subspace of the environment tensor space. The bounds for the values of $A$ and $B$ for the case of general set of unitaries are found by calculating the singular values of the {\it analysis operator} $T: V \rightarrow \mathbb{C}^n$, defined as the linear transformation that takes a vector to its projections on the frame elements
\begin{align}
    T(v) = \left\{ ({\mathbf{e}}_k, v)\right\}_{k=1...n}.
\end{align}
The largest and smallest singular values of $T$ correspond to the $A$ and $B$ bounds. In the case of projection on unitary channel, the maximal magnitude of the coordinate vector is reached for $E = \pi_I$, which gives the maximal projection $(\pi_U,\pi_I) = 1$ and the upper bound $B = n$. The lower bound $A$ is determined by the lowest eigenvalue of $T$, which depends specifically on the choice of the unitary projectors that construct the linear frame. This is the more interesting of the two bounds, because it is directly related to the reconstruction accuracy of the environment tensor. 
To recover the environment tensor from its linear frame representation we use the {\it dual frame} $\{{\Tilde{\mathbf{e}}}_k\}$, a set of vectors defined by their inversion property
\begin{equation}
    v = \sum{(\Tilde{\mathbf{e}}_k, v) \mathbf{e}_k}
\end{equation}

The dual frame is particularly useful in the case of environment tomography, because they are the tool which enables us to recover the original environment tensor out of its projections. 
The dual frame can be found explicitly by constructing the {\it frame operator} $S = T^* T$, where  $T^*$ is the adjoint to the analysis operator- $ T^*(\{c_k\}) = \sum{c_k \mathbf{e}_k}.$. Using the frame operator we can construct the dual-frame vectors by the following formula:
\begin{align}
    \Tilde{\mathbf{e}}_k = S^{-1} {\mathbf{e}}_k =  \left(T^* T\right)^{-1} {\mathbf{e}}_k 
\end{align}

This formula is nearly identical to the solution of linear regression, and the matrix representation of the inverse of the frame operator appears many times in the linear-inversion based formalism in section \ref{sec:regression} of the main text as $M^\dagger M$ (see Eq. \ref{eq:inv_solution} for an example). For a closer inspection of $M^\dagger M$ in the case of 2-unitary design gate set see section \ref{app:opti_2_design}.

Lastly, a {\it tight frame} is defined as a frame in which $A = B$. This is the case when the frame is completely unbiased, and does not overlap with any vector more than others. In our case, a gate-set distributed according to the Haar-random unitary distribution is {\it nearly} tight, as it does prefer the constant tensor term and probe it more strongly. If we ignore this term by looking at the subspace of all environment tensors with zero constant term (and have an average term of zero), then a tight frame is and sufficient and necessary for both 2-unitary design, and optimal tomography accuracy. 
 
\section{The horizontal decomposition and the non-measurable degrees of freedom \label{app:decomp}}

To analyze the task of environment tomography we first present several decomposition of the environment tensor in different bases.

The horizontal decomposition rewrites the environment tensor as a sum of many tensor products of two separate Pauli strings
\begin{align}
    \Tderivderiv{E} = \sum_{i,j} e_{i,j} \ABUVdiagClean{P_i}{P_j}
\end{align}
In this decomposition the Pauli tensors connect between the left and the right side of the diagram, and the reduced cost function take the form of a tensors chained together in a loop:
\begin{align}
    \Tdiag{U}{U} &= \sum_{i,j} e_{i,j} \ABUVdiag{P_i}{P_j}{U}{U} \\
    &= \sum_{i,j} e_{i,j}\Tr(\sigma_i U \sigma_j U^\dagger)
\end{align}
The horizontal base is orthogonal under contraction
\begin{align}
    \ABUVdiagClean{P_i}{P_j} \circ \ABUVdiagClean{P_{i'}}{P_{j'}} &= \PiPj{P_i}{P_j}{P_{i'}}{P_{j'}}\\
    &= \tr(P_i P_{i'})\tr(P_j P_{j'}) = \delta_{i,i'}\delta_{j,j'} \nonumber
\end{align}
and so each basis element represent an independent coordinate of the environment tensor.
We find this basis useful later, both for discerning the measurable and non-measurable components of $E$, and for constructing a set of gates that will perform full tomography on environment tensor.

Similarly, it is possible to define a vertical basis, in which the environment is decomposed into sum of tensor products of left environment and right environment tensor. Taking the left and right environments to be Pauli strings creates a complete orthogonal base.

After examining the various components of the environment tensor in the horizontal basis, we can distinguish between the elements that contribute to the reduced cost functions and those that do not.
Particularly, for n-qubit gate, the basis elements with either $P_i$ or $P_j$ set to the identity string has no effect on the reduced cost function, as can be easily checked
\begin{align}
    \ABUVdiag{I}{P_j}{U}{U} = \Tr(I U P_j U^\dagger) = \Tr(P_j) = 0
\end{align}
which is true for every non-identity Pauli string $P_j$.

Conversely, due to the orthogonality of the base elements, all the rest of the elements have a unique impact on the reduced cost function, in the sense that each element cannot be expressed as linear combination of the rest  and each element has some gates which yield non-zero values. The gates which get non-zero contribution from an $(i,j)$ environment component can be found by simple diagonalization of both Pauli strings
\begin{align}
    \Tr( P_i U P_j U^\dagger) &= \Tr( (V_i \Lambda V_i^\dagger) U (\Lambda V_j^\dagger) U^\dagger) \nonumber \\
    &= \Tr( \Lambda (V_i^\dagger U V_j) \Lambda (V_j U V_j^\dagger)^\dagger)
\end{align}
when $\Lambda$ is the diagonal matrix of the eignevalues ($\Lambda=Z\otimes I^{\otimes n-1}$), and $V_i$ are the diagonalizing transformation matrices . Choosing $U = V_i V_j^\dagger$ will evidently yield $\Tr( P_i U P_j U^\dagger) = 2^n$, which is the maximal value for a single matrix substitution.

From the properties above, we can count the number of independent measurable components in the environment of an $n$-qubit gate. Having $4^n - 1$ non-identity Pauli strings and taking into account the additional constant component of $P_i = P_j = I$, we get
\begin{align}
    N_{\rm measurable} &= (4^n -1)^2 + 1 \\
                        &= 4^{2n} -2\cdot 4^n + 2,
\end{align}
which is a dimension independent on the choice of basis for environment component.

\section{Linear cost functions tomography}
\label{app:Linear environment tomography}

In several variational algorithms the cost function can be expressed as a perfect square of a linear function. 
This is the case, for example, in fidelity estimation and optimization for optimal state encoding\cite{bendov2024approximate,ran2020encoding,rudolph2023decomposition} and circuit recompilation\cite{khatri2019quantum,heya2018variational}, where the cost function can be expressed in terms of an overlap between two states.
In these cases, the environment of the reduced cost function can be described using fewer degrees of freedom and the tomography algorithm can be reduced into a simpler set of measurements. 

The quadratic cost function can be rewritten in terms of a linear overlap function
\begin{align}
    f(U,U)  &= \abs{\tr(E_{\rm L}^\dagger U)}^2= \LinearEnvdiag{U}{U}
\end{align}
where the linear environment $E_{\rm L}$ is defined up to a global phase.
To extract the linear environment we need to find a set of measurement that will probe all the elements of the gradient. 

One way of constructing a tomography measurement set is by expressing the linear environment tensor in the Pauli basis as a sum of Pauli strings $E_{\rm L}=\sum_i{{E_{\rm L,i}}\sigma_{i_1}\sigma_{i_2}....\sigma_{i_n}}$, and to collect gates which probe the amplitudes and phases of the different components. 
Substitution of Pauli strings can successfully retrieve the norm of the different gates, while the relative phases can be probed using a combination of Pauli strings. 
\begin{align}
    &\mathcal{P}_{\rm Lin} = \{\sigma_{i_1}\sigma_{i_2}\ldots\sigma_{i_n} \mid i_k \in \{0,\ldots,3\}\} \\
    &\mathcal{T}_{\rm Lin} = \left\{\frac{1}{\sqrt{2}}\left(\sigma_{i_1}\ldots\sigma_{i_n} + \zeta \sigma_{j_1}\ldots\sigma_{j_n}\right) \middle| 
    \begin{array}{l}
      i_k \in \{0,\ldots,3\} \\
      \zeta = \pm 1, \pm i
    \end{array}\right\} \nonumber 
\end{align}

In order to make the calculation possible on a real quantum device, the combination of Pauli strings are chosen such that their sum is a unitary gate. The phase $\zeta$ is chosen to be $\pm 1$ for anti-commuting Pauli strings, and $\pm i$ for combinations of commuting Pauli strings.

Using the unitary Pauli combinations above we can extract all the components of the environment tensor- $E_{\rm L,i}$, up to a global phase. To perform a complete tomography we measure the circuit with substitutions of $4^k$ Pauli strings for amplitude extraction by $f(P_i) = \abs{{E_{\rm L,i}}}^2$, and $4^k - 1$ Pauli combinations for relative phases measurement $f(T_{ij}) = \abs{{E_{\rm L,i}}}^2 + \abs{{E_{\rm L,j}}}^2 +2\re{\zeta {E_{\rm L,i}} {E_{\rm L,j}}}$. 
There is a freedom of choice for the combinations used to extract all the relative phases. We can use individual measurement of relative phases to link the phases of different amplitudes, which build up to either a linear chain of relative phases (which takes $4^k-1$ measurements), or to other forms of fully connected graphs. 
To make the tomography easier to implement, the Pauli combinations used for phase extraction can be chosen by combining strings that differ only on one site, resulting in a fully-separable gate consisting of tensor product of single-qubit operations. This make the case of environment tomography for pure-square cost function much easier to implement on real hardware, even for many-body quantum system, as the gate ensemble used for tomography is local in this particular case.








\section{Regression-based tomography \label{app:reg}}


\subsection{Basis free formulation of the linear inverstion tomography}
In section \ref{sec:regression} we have introduced the linear equation system that connect between measurements results and the environment tensor, and described how we can invert it to recover the tensor out of measurements. We have described the  matrix $\hat{M}$ and the vector of amplitudes $\vec{a}$ using a general basis to the linear space of environment tensors $\left\{B_i\right\}$. The linear system itself though is independant of basis, which invites us to describe it in basis-free formulation using operators and superoperators, which we describe here. 

We start by rewriting eq. \ref{eq:linear_reg} by treating $a_E$ as the environment tensor itself and $M$ as the projection tensor onto a single measurement shadow
\begin{align}
    \UVSpiderDiagramAnnotated{U_i}{U_i^\dagger} \circ \Tderivderiv{E} &= f_{\rm single-shot}(U_i)    
\end{align}
when the inner dimensions of $E$ and $M$ are summed over by contraction. Multiplying both sides by $\hat{M}^\dagger$ gives:
\begin{align}
    \frac{1}{N}\sum_i{\left\{\UVSpiderDiagramAnnotated{U_i}{U_i^\dagger}\UVSpiderDiagramAnnotated{U_i^\dagger}{U_i}\right\}} \circ \Tderivderiv{E} &= \frac{1}{N} \sum_i{\left\{f_{\rm single-shot}(U_i) \UVSpiderDiagramAnnotated{U_i}{U_i^\dagger}\right\}} \\   
    \frac{1}{N}\sum_i{\left\{\UVSpiderDiagramAnnotated{U_i^\dagger}{U_i}\Tdiag{U}{U}\right\}} &= \frac{1}{N}\sum_i{\left\{f_{\rm single-shot}(U_i) \UVSpiderDiagramAnnotated{U_i^\dagger}{U_i}\right\}}      \\
    \hat{\bm{O }}(E) &=  \underset{U\in \mathcal{U}_s}{\mathbb{E}}\left[ \mathcal{S}_\text{single-shot} \right] 
\end{align}
where $\hat{\bm{O }}= \frac{1}{N}\sum_i{\left\{\scalebox{0.8}{\UVSpiderDiagramAnnotated{U_i}{U_i^\dagger}\UVSpiderDiagramAnnotated{U_i^\dagger}{U_i}}\right\}}$ is the second moment tensor that acts as a function- transforming one environment tensor to another by set of projections, and on the right hand size the sum of measured shadows from Eq.~\ref{eq:single_shot_shadow}.
To obtain the corrected environment tensor, we have to apply the inverse transformation of $\hat{\bm{O }}^{-1}$ by matrix inversion, leading to Eq.~\ref{eq:inv_solution} in the main text.

\subsection{Derivation of the reconstructed error formula, in Eq.~\ref{eq:linear_bound} of main text \label{app:linear_bound_proof}}
To calculate the expected deviation of the reconstructed environment tensor from the original one we calculate the expected distance between the extracted vector of amplitude $\vec{v}_{\rm op}$ and the vector of original amplitudes $\vec{E}$. 

\begin{align}
    {\rm Var}(\vec{v}_{\rm op}) = \mathbb{E}\left[\norm{\vec{v}_{\rm op} - \vec{E}}^2\right] &= \mathbb{E}\left[\norm{(\hat{M}^\dagger \hat{M})^{-1}\hat{M}^\dagger\delta_{\vec{\phi}_s} }^2\right] \\
    &= \mathbb{E}\left[\vec{\delta}_{\vec{\phi}_s}^\dagger \hat{M} (\hat{M}^\dagger \hat{M})^{-2} \hat{M}^\dagger \vec{\delta}_{\vec{\phi}_s} \right] \\
    &= \sum_{i,j}{ (\hat{M} (\hat{M}^\dagger \hat{M})^{-2} \hat{M}^\dagger)_{ij} \mathbb{E}\left[\left({{\delta}_{\vec{\phi}_s}}\right)_i\left({{\delta}_{\vec{\phi}_s}}\right)_j\right]}
\end{align}
Because elements of $\phi_s$ originate from separate independent shots, correlation between different indices is zero-  $\mathbb{E}\left[\left({\delta}_{\vec{\phi}_s}\right)_i\left({\delta}_{\vec{\phi}_s}\right)_j\right] = {\rm Var}\left(\left({\delta}_{\vec{\phi}_s}\right)_i\right)\delta_{i,j}$
\begin{align}
    {\rm Var}(\vec{v}_{\rm op}) &= \sum_{i,j}{ (\hat{M} (\hat{M}^\dagger \hat{M})^{-2} \hat{M}^\dagger)_{ij} \mathbb{E}\left[\left({\delta}_{\vec{\phi}_s}\right)_i\left({\delta}_{\vec{\phi}_s}\right)_j\right]} \\
    &= \sum_{i}{ (\hat{M} (\hat{M}^\dagger \hat{M})^{-2} \hat{M}^\dagger)_{ii} {\rm Var}({\phi}_{s,i})}
\end{align}
Assuming that the choices of unitaries are not correlated with the variance of the cost function samples, we can approximate the sum as 
\begin{align}
    {\rm Var}(\vec{v}_{\rm op}) &\approx \frac{1}{N} \sum_{i}{ (\hat{M} (\hat{M}^\dagger \hat{M})^{-2} \hat{M}^\dagger)_{ii}}\sum_i{{\rm Var}({\phi}_{s,i})} \\
    &= \frac{1}{N} {\rm Tr}\left(\left(\frac{\hat{M}^\dagger \hat{M}}{N}\right)^{-1}\right) \Bar{{\rm Var}({\phi}_{s})}
\end{align}
\subsection{Optimality of 2-unitary design \label{app:opti_2_design}}
In this section we give a proof for the optimality of 2-unitary design for environment tensor tomography. This result is equivalent to the optimality of 2-unitary design for quantum channel tomography in ref. \cite{scott2008optimizing}.
To check the performance of 2-unitary design we look at the singular values of the second moment matrix $M^\dagger M$
The elements of the second moment matrix are :
\begin{align}
    (M^\dagger M)_{i,j} = \sum_{U\in \mathcal{U}}{\TdiagB{B_i}{U}{U}\;\TdiagB{B_j}{U}{U}}
\end{align}
The trace of $M^\dagger M$ is independent of the choice of gates
\begin{align}
    {\rm Tr}(M^\dagger M) = \sum_i{\abs{{\rm tr}\left(U_i^\dagger U_i\right)}^2} = N d^2
\end{align}
with $d$ being the size of the unitary matrix $U$.
For a gate set $\mathcal{U}$ that acts as a 2-unitary design, the summation over unitaries can be replaced with an integral over the uniform Haar distribution. Solving second-order Haar-random integrals can be done analytically using Weingarten calculus \cite{kostenberger2021weingarten, mele2024introduction}
\begin{align}
    \frac{1}{N}(M^\dagger M)_{i,j} &= \int_{\rm Haar}{\TdiagB{B_i}{U}{U}\;\TdiagB{B_j}{U}{U} \;dU}\\
                        &= \frac{1}{d^2-1} \left( \vcenter{\hbox{\includegraphics[scale = 0.4]{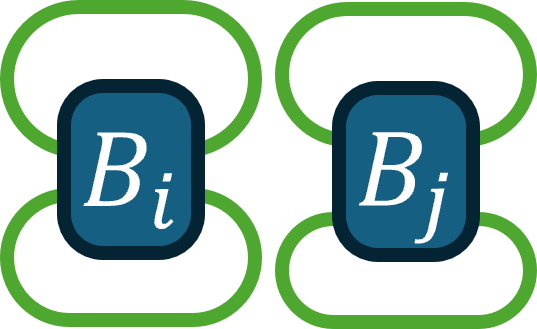}}}+\vcenter{\hbox{\includegraphics[scale = 0.4]{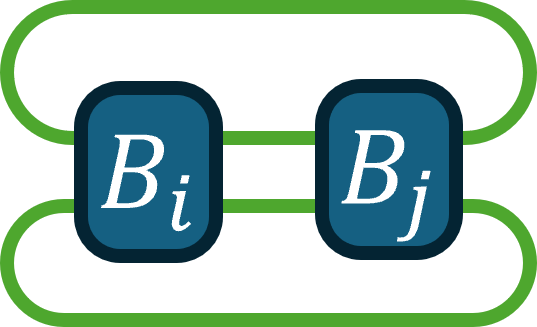}}} \right) -  \frac{1}{d(d^2-1)}\left(\vcenter{\hbox{\includegraphics[scale = 0.4]{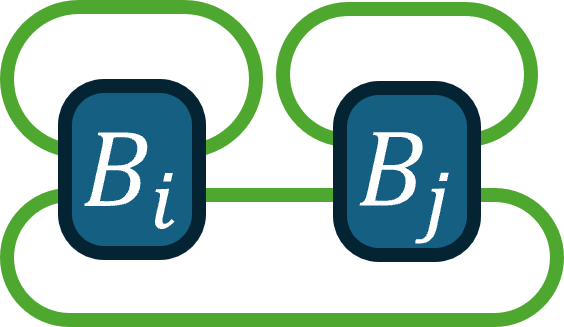}}} + \vcenter{\hbox{\includegraphics[scale = 0.4]{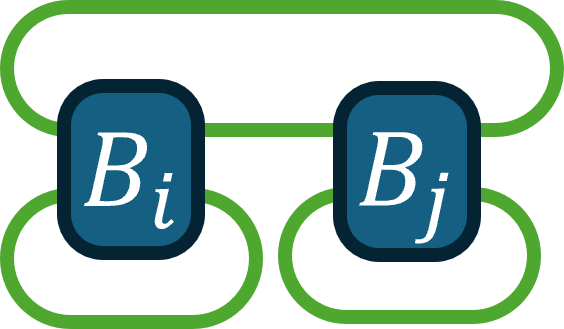}}}\right) \\
                        &= \frac{1}{d^2-1} (d^2\delta_{i1}\delta_{j1}  + \delta_{ij} ) - \frac{1}{d(d^2-1)}(d\delta_{i1}\delta_{j1} + d\delta_{i1}\delta_{j1})\\
                        &= \begin{cases}
                            1 &\text{$i=j=1$}\\
                            \frac{1}{d^2 - 1} &\text{$i=j>1$} 
                            \end{cases}
\end{align}
Here we used the horizontal Pauli basis for the relevant subspace of the environment tensor, where $B_1$ is the constant background component. The value for each of the tensor diagrams can be explained as by the rules of traces and contractions between Pauli string- the trace of any non-identity Pauli string and the contraction between any non-identical Paulis is zero. In the evaluation of the diagrame we neglected the non-measurable Pauli elements that have identity on the upper string and non-identity Pauli string on the bottom string, or vise versa. 

The value of $(M^\dagger M )_{11}$ is independent on the choice of gate set and always averages out to 1.  In the same manner the constant basis element $B_1$ is an eigenvector of $M^\dagger M$, and so $s_1 = 1$ for any choice of gate set.
According to Eq. \ref{eq:linear_bound}, given no prior information on the environment tenosr, the expected variance of the reconstructed environment tensor is proportional to
\begin{align}
    {\rm Var}(\Vec{a}) \propto {\rm Tr}\left(\left(\frac{M^\dagger M}{N}\right)^{-1}\right) = \sum_i{\frac{1}{s_i}} = 1 + \sum_{i=2}^{n_b}{\frac{1}{s_i}}
\end{align}
Given that all the sum of all singular values is predetermined according to the matrix trace, finding the minimal configuration of eigenvalues is a simple constraint optimization problem, and it is easy to show that the minimum configuration is the uniform partition, as in the case of 2-unitary design.
In the other direction, if a set of gates is given so that the second moment matrix has the configuration of unitaries described above, then 
\begin{align}
    {\rm Tr}\left(\left(\frac{M^\dagger M}{N}\right)^2\right) = \sum_i{s_i^2} = 1 + \sum_{i=2}^{n_b}{\frac{1}{(d^2-1)^2}} = 2
\end{align}
But at the same time, due to the cyclic nature of the trace function, the trace is equal to the average fidelity squared between any pair of matrices
\begin{align}
    {\rm Tr}\left(\left(\frac{M^\dagger M}{N}\right)^2\right) = {\rm Tr}\left(\left(\frac{M M^\dagger}{N}\right)^2\right) = \frac{1}{N^2}\sum_i{\abs{tr(U_i^\dagger U_j)}^4} = \mathcal{P}(\mathcal{U})
\end{align}
which is knows as the {\it frame potential} of set of gates. It has been proven \cite{gross2007evenly} that the minimal frame potential is a sufficient and nessesary condition for a 2-unitary design- $\mathcal{P}(\mathcal{U}) = 2$ iff the set $\mathcal{U}$ is a unitary design, which completes the proof.
\section{Uniform landscape tomography \label{app:shadow}}
\subsection{Environment reconstruction by Haar integration over shadows}
The equation for the reconstructed environment using shadow tomography (Eq. \ref{eq:Shadow_tomography}) demonstrate the optimal bound for tomography using a minimal number of shots as discussed in \ref{app:Linear environment tomography}. To derive the expression for the averaged unitary shadows we can use the linear inversion formalism with a set of uniformly sampled gates. 
Here we show a direct derivation of the formula which uses integration over Haar random unitaries:
\begin{align}
    \int_{\text Haar}{\mathcal{S}_{\rm exact}(U)dU} &= \int_{\text Haar}{\left(\Tdiag{U}{U}\right) \UVSpiderDiagram dU}\\
    &= \frac{1}{d^2 - 1}\left(\vcenter{\hbox{\includegraphics[scale = 0.4]{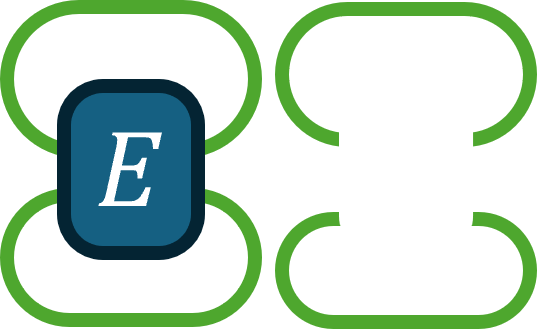}}} + \vcenter{\hbox{\includegraphics[scale = 0.4]{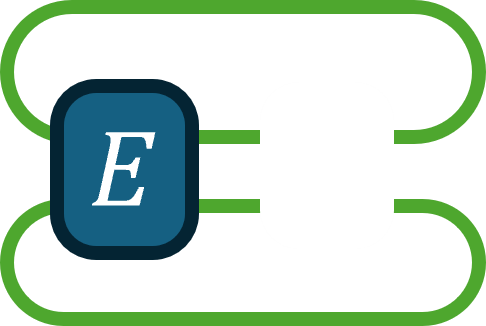}}}\right) - \frac{1}{d(d^2 - 1)}\left(\vcenter{\hbox{\includegraphics[scale = 0.4]{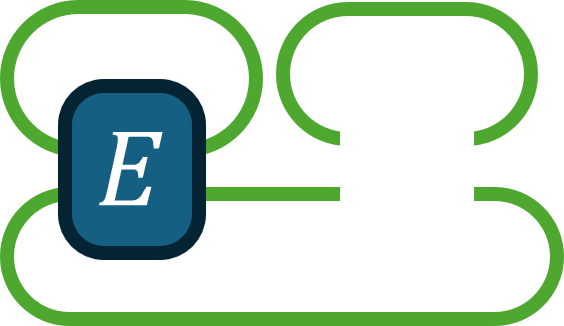}}} + \vcenter{\hbox{\includegraphics[scale = 0.4]{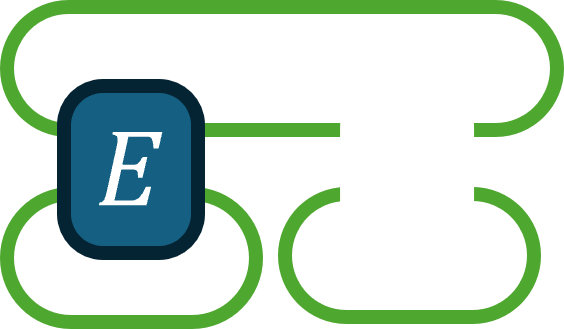}}}\right)
\end{align}

The first two elements contribute a constant term and a term of the original environment tensor. To make sense of the 2 last terms, we check their behavior under the contraction of a gate $V$ and $V^\dagger$:
\begin{align}
    &V\circ \left(\int_{\text Haar}{\mathcal{S}_{\rm exact}(U)dU} \right)  \circ V^\dagger \\
    &= \frac{1}{d^2 - 1}\left(\vcenter{\hbox{\includegraphics[scale = 0.4]{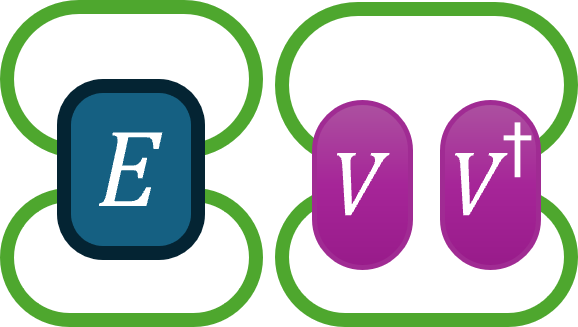}}} + \vcenter{\hbox{\includegraphics[scale = 0.4]{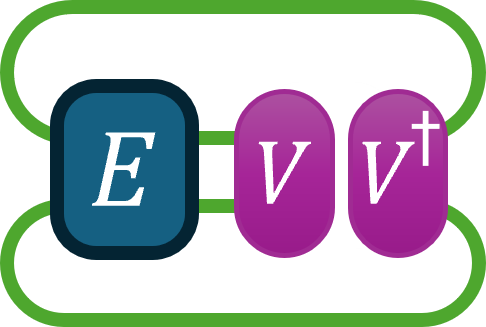}}}\right) - \frac{1}{d(d^2 - 1)}\left(\vcenter{\hbox{\includegraphics[scale = 0.4]{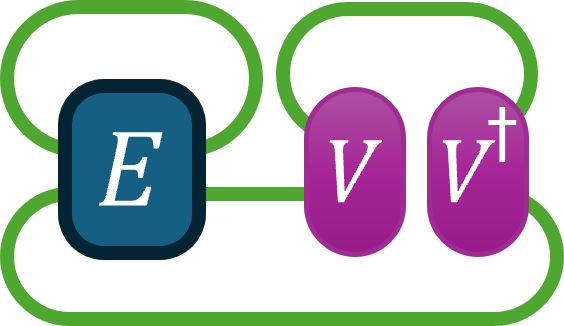}}} + \vcenter{\hbox{\includegraphics[scale = 0.4]{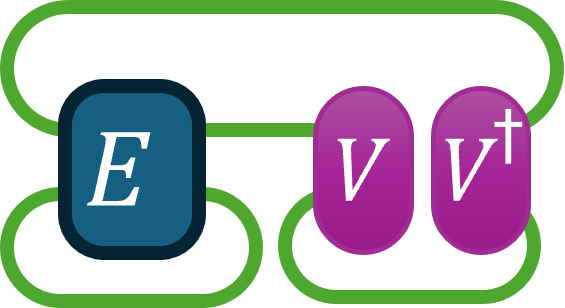}}}\right)
\end{align}
Disregarding terms that may appear only in non-symmetric unitary contraction, the last two terms add a constant value to the environment reconstruction, and can be summed up with the constant background term to get the final form of the environment tensor reconstruction:


\begin{align}
    \int_{\text Haar}{\mathcal{S}_{\rm exact}(U)dU} &= \left(\frac{1}{d^2 - 1} - \frac{2}{d^2(d^2 - 1)}\right) \left(\vcenter{\hbox{\includegraphics[scale = 0.4]{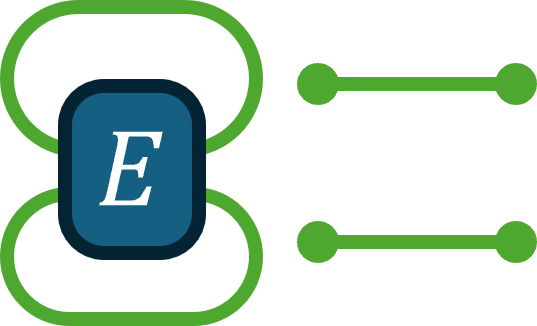}}}\right) + \frac{1}{d^2 - 1}\vcenter{\hbox{\includegraphics[scale = 0.4]{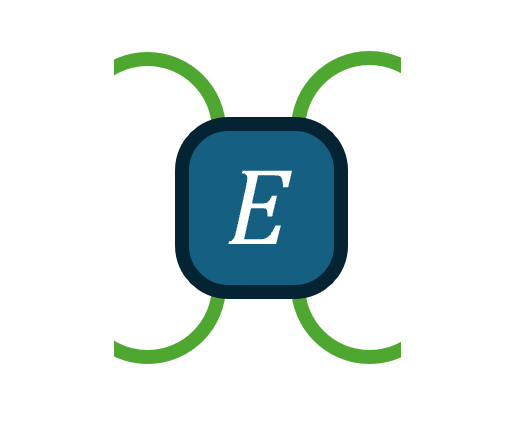}}} \\
    &= \frac{d^2 - 2}{d^2(d^2 - 1)} \left(\vcenter{\hbox{\includegraphics[scale = 0.4]{diagrams/Picture5.png}}}\right) + \frac{1}{d^2 - 1}\vcenter{\hbox{\includegraphics[scale = 0.4]{diagrams/Picture6.png}}}
\end{align}

which completes the derivation.

\subsection{Connection between regression and shadow tomography}

The results for environment tensor reconstruction using shadow tomography can also be derived using the formulation of linear inversion-based tomography. Recall that the second moment matrix $M^\dagger M$ for a 2-unitary design gate set is a diagonal matrix of the form:
\begin{align}
    (M^\dagger M)_{ii} = \begin{cases}
                            1 &\text{$i=1$}\\
                            \frac{1}{d^2 - 1} &\text{$2\leq i\leq (d^2 - 1)^2 + 1$} 
                            \end{cases}
\end{align}
when $i = 1$ correspond to the constant background term and the terms form a basis to the relevant subspace of the environment tensor. Applying the matrix on the vectors of amplitude $\vec{v}_E$  gives different weight to the constant background compared to the other terms
\begin{align}
    \mathbb{E}(\mathcal{S}_{\text{single-shot}}) &= \sum_i{(M^\dagger M\vec{v}_E)_i \Tderivderiv{B_i}} \\
                                                &= \sum_i{(M^\dagger M)_{ii} (E\circ B_i) \; \Tderivderiv{B_i}}\\
                                                &= (M^\dagger M)_{11} \frac{{\rm tr}(E)}{d} \; \frac{1}{d}\ParallelLines{-0.2}{0.2} + (M^\dagger M)_{ii} \left(\Tderivderiv{E} - \frac{{\rm tr}(E)}{d} \;\frac{1}{d}\ParallelLines{-0.2}{0.2}\right)\\
                                                &= ((M^\dagger M)_{11} - (M^\dagger M)_{ii})\frac{{\rm tr}(E)}{d} \; \frac{1}{d}\ParallelLines{-0.2}{0.2} + (M^\dagger M)_{ii} \Tderivderiv{E} \\
                                                &= \frac{d^2-2}{d^2(d^2-1)} {\rm tr}(E) \ParallelLines{-0.2}{0.2} + \frac{1}{d^2 - 1} \Tderivderiv{E}
\end{align}
\section{Stabilizer-based tomography \label{app:stab}}

\subsection{Greedy algorithm for searching minimal cover}
In the main text we have described a way to construct a tomography gate set by combining separate sets of $2^{2k}$ gates. Each set is generated by a single Clifford gate and fully characterize a subset of $2^{2k}$ environment components. For a full characterization of the environment tensor, these subsets have to be chosen and combined in a way that covers all components of the environment tensor with minimal overlap. For the simple case of 1-qubit gate the optimal solution can be found easily - 3 sets of four gates each, which function also as a unitary 2-design for single qubit gates.
In the general case it is harder to find the minimal solution, and to search for a good configurations we have employed a greedy search algorithm that try to minimize the overlap for every addition of a new subset. 

The algorithm starts with an empty gate set and an empty set of measured components, and at each step go over a pool of candidate Clifford gates and their matching subsets and rank them according to their overlap with components that have been already measured. Then the algorithm picks randomly one of the sets with minimal overlap and append it to the set of tomography gates. The algorithm finishes when all components have been covered. We run the algorithm multiple times, and take the result with the least amount of subsets.

Using the brute-search algorithm we have found a cover set for 2-qubit gates which contain 17 group of gates- with a total of $272$ gates The list of 17 gates, up to pauli-string multiplication is given by:

\begin{align}
&(S^{\dagger}X\otimes S^{\dagger} )\;\;{\rm CNOT}\;\;(SHS^{\dagger}Y\otimes HZ)\;\;{\rm CNOT}\;\;(HX\otimes HSHX) \\
&(HSHY\otimes S^{\dagger}Y)\;\;{\rm CNOT}\;\;(HS^{\dagger}Z\otimes HX) \\
&(HSZ\otimes SX)\;\;{\rm CNOT}\;\;(HZ\otimes SHZ)\;\;{\rm CNOT}\;\;( X\otimes HS^{\dagger}Y) \\
&(HS^{\dagger}X\otimes HSY) \\
&(HZ\otimes SHZ)\;\;{\rm CNOT}\;\;(HSZ\otimes SHS^{\dagger}X) \\
&(HY\otimes  Y)\;\;{\rm CNOT}\;\;(S^{\dagger}HZ\otimes SHX) \\
&(S^{\dagger}HZ\otimes  Z)\;\;{\rm CNOT}\;\;( Y\otimes HSHZ) \\
&(SY\otimes HZ)\;\;{\rm CNOT}\;\;(HX\otimes HSX) \\
&(SY\otimes S^{\dagger}Z)\;\;{\rm CNOT}\;\;( Y\otimes S ) \\
&(S^{\dagger}HX\otimes HZ)\;\;{\rm CNOT}\;\;(S^{\dagger}HX\otimes S^{\dagger}X)\;\;{\rm CNOT}\;\;(SX\otimes S^{\dagger}Z) \\
&(SX\otimes S^{\dagger}HX)\;\;{\rm CNOT}\;\;(HSHZ\otimes HZ)\;\;{\rm CNOT}\;\;(SHY\otimes SZ) \\
&( Y\otimes HSHY)\;\;{\rm CNOT}\;\;(SHZ\otimes SY)\;\;{\rm CNOT}\;\;(HY\otimes HSHX) \\
&(HY\otimes HSY)\;\;{\rm CNOT}\;\;(HS^{\dagger}X\otimes H )\;\;{\rm CNOT}\;\;(S^{\dagger}X\otimes SHS^{\dagger}Y) \\
&(HS^{\dagger}Y\otimes HSY)\;\;{\rm CNOT}\;\;(SHX\otimes HY)\;\;{\rm CNOT}\;\;(S^{\dagger}Y\otimes S^{\dagger}Z) \\
&(HX\otimes HS^{\dagger} )\;\;{\rm CNOT}\;\;(H \otimes SY)\;\;{\rm CNOT}\;\;(HSHX\otimes SY) \\
&( Z\otimes HSHY)\;\;{\rm CNOT}\;\;(SHY\otimes HX)\;\;{\rm CNOT}\;\;( X\otimes S^{\dagger}Y) \\
&(HZ\otimes HSX)\;\;{\rm CNOT}\;\;(S^{\dagger}H \otimes S^{\dagger}HY)\;\;{\rm CNOT}\;\;(HSZ\otimes  I)
\end{align}

Incidentally, this set of gate contain only gates with up to $2$ 
$\rm CNOTs$, with an average of $26/17 \approx 1.588$ application of CNOT gates per single gate. 
As mentioned in the main text in section \ref{sec:regression}, the efficiency of our designed gate set is close to optimal, with only 5.8\% increase in the number of shots over using an optimal 2-unitary design.

From further brute-force search over all possible cover sets, we ruled out the existence of any smaller informationally-complete Clifford gate set with the structure of 16 gates per group, and searches for similar sets without grouping of 16 gates has not been successfull.


\subsection{Minimizing CNOT count for full Tableau-based tomography \label{app:CNOTCount}}

Given an environment tensor of a general k-qubit gate, it is natural to wonder about the minimal resources that are necessary to recover all the information of the cost function.  
The minimal number of CNOT gates is important for assessing the accuracy of real-life implementation of the tomography algorithm on quantum hardware. 
Looking at the tomography task from the perspective of the Pauli basis, each component of the environment tensor $Pi_, P_j$ can be probed by finding a Clifford gate that follow the transformation $U^\dagger P_i U = P_j$, as described in section \ref{sec:stab} of the main text.

Going over all pairs of Pauli string, we can describe an efficient k-qubit circuit that transform one Pauli string to the other using the generator of the Clifford group- the 1-qubit gates $S$ and $H$ and the 2-qubit gate $\rm CNOT$. Each gate apply a transformation on the Pauli string in the Heisenberg picture, and by identifying the unique role of the CNOT gate as a multi-qubit transformation we can design the circuit that require the minimal number of CNOT gates for the transformation. Here we describe a construction of a Clifford circuits that, given a general environment tensor, probe all the relevant components using the minimal number of CNOT gates. We go over all components in the Pauli basis and describe the minimal circuit that can probe it.

First we consider the case of using only 1-qubit gates and excluding CNOT entirely. The possible transformation then only act on each single qubit separately- 
\begin{align}
    U^\dagger P_i U = (U_1^\dagger P^1_i U_1) \otimes (U_2^\dagger P^2_i U_2)  \otimes ... \otimes (U_k^\dagger P_i^k U_k) = P_j
\end{align}
where $P^l_i$ is the l-th Pauli gate of the Pauli string $P_i$. Given that each Pauli gate undergo a unitary transformation independently, the possible transformations consist of combinations of the 10 possible single-qubit gate pairing: 9 Pauli combinations - $\sigma_i \rightarrow \sigma_j$ for $i,j = 1,2,3$ with an additional transformation on the identity $I \rightarrow I$. Note that the identity cannot transform to non-identity Pauli gate and vice-versa, a rule which apply to all single-qubit transformation independently.
For example, given no CNOT gates, the following 5-qubit transformation is allowed
\begin{align}
    XIXZ \rightarrow YIZZ,
\end{align}
generated for example by $U = C_{\rm up} \otimes I \otimes C_{\rm down} \otimes I$, 
whereas the transformation
\begin{align}
    XIXZ \rightarrow XZZI
    \label{eq:misaligned_paulis}
\end{align} 
cannot be implemented as the identity gates are not aligned between the different qubit. Adding CNOT gates to the circuit introduces a way to transform identity gate into a Pauli gate using a second ancillary qubit. As presented in Eq. \ref{eq:CNOT_transformation}, the CNOT gate induces the transformations $X I \leftrightarrow X X$ and $I Z \leftrightarrow Z Z$, and given an identity operator alongside additional Pauli operator acting on an ancilla qubit we can apply local transformations followed by a CNOT gate to cancel the identity operator after transformation. The number of CNOT gates determine the number of possible identity operator cancellations, which are required in case of non-aligned identity operators. 
Here the number of CNOT gates depends on the connectivity between the different qubits, and here we consider two possibilities - all to all connectivity and 1-D geometry with open boundaries.

For all to all connectivity, a general transformation can be constructed by identifying all misaligned identity operators, and applying CNOT gates between each one of them and an ancillary qubit with non-identity Pauli operator. 
For example, taking the transformation of Eq. \ref{eq:misaligned_paulis}, we can apply the following transformations- $U = {\rm CNOT}_{23}{\rm CNOT}_{43}H_3$.
The first CNOT gate ${\rm CNOT}_{43}$ transform $Z_3 Z_4$ into $Z_3 I$, matching the target Pauli string, and  ${\rm CNOT}_{23}$ transform $I_2 Z_3$ into $Z_2 Z_3$.
The maximal number of CNOT gates for all to all connectivity is given by the number of maximal non-matching pairs- $k$. 

For the case of linear $1D$ connectivity the qubits are aligned in an array and CNOT gates are only possible between adjacent qubits. In this case some transformations will require additional CNOT gates when neighboring qubits have pairs of identity operators and a single local CNOT gate is unable to cancel unmatched identity terms. 
For example we consider the transformation $XII \rightarrow IIX$. Here we need to transform $X_1$ into $I_1$ and $I_3$ into $X_3$, but applying CNOT on either the 1-2 and the 2-3 pair will not induce the correct transformation. If we allow to perform a CNOT gate between 1-3, then the minimal circuit would be $U = {\rm CNOT}_{13}{\rm CNOT}_{31}$. Otherwise, the best strategy is to transform the gates

\section{Environment tomography for constraint gate with limited CNOT count\label{app:CNOTRestrict}}

Given a constraint on the number of CNOT applications for a k-qubit gate, it is possible to re-count the number of elements in the horizontal Pauli basis which can be independently probed. This combinatorial problem produce the expressions in Eq. \ref{eq:no_meas0} and \ref{eq:no_meas} of the main text.


We start with counting the number of components which are measurable without using CNOT gates entirely. These components has been mentioned in section \ref{app:CNOTCount}, and are the elements that include no unpaired identity operator on the same site. The possible pairings in this case are either two non-identity Pauli strings or two identity operators, with a total of 10 possible pairing per single qubit. For a k-qubit gate we get that the total number of measurable elements is $N_{\rm measurable}^{{N_{\rm CNOT} = 0}} = 10^k$.

Now we again introduce CNOT gates - which enables the probing of components with unpaired identities by using other pairs as ancillas of 2-qubit operations. 
Let us consider the case of a single CNOT gate - we can then allow a single unpaired identity, which adds 6 possible pairs (identity with the 3 Pauli gates, multiplied two possible placements of the identity). There is a single exception that we need to exclude- in the case where all the rest of the  pairs are identities we cannot correct the unpaired identity using CNOT gate. Counting all possible configurations, we have $k$ possible positions for 6 possible pairs of unmatched identities. The rest of the pairs having 10 possible pairings - with the exception of uniform identity pairs. Combining it all into a single formula:
\begin{align}
    N_{\rm measurable}^{{N_{\rm CNOT} = 1}} = 10^k + 6k(10^{k-1} - 1)
\end{align}

With two or more CNOT gates, the picture is similar. With $l$ CNOT gates we can have $l$ pairs with unmatched identities that can be dealt with using proper placement of CNOT gates. Counting the number of components with $l$ pairs, we get $6^l$ unpaired identity combinations placed at ${\binom{k}{l}}$ possible position combinations.
The rest of the pairs can have 10 possible pairing with a single exception when all pairs are identity. This last exception however is only relevant to a fraction of the cases -when all unmatched pairs are aligned together with all identities placed at the upper of lower Pauli string together, otherwise it is possible to eliminate all unmatched pairs with CNOT application between them without needing an external ancilla. As a result we subtract from the general $10^{k-l}$ possible pairing a non-integer number which exclude the exception only for fully aligned unmatched identity pairs- $10^{k-l} - 0.5^{l-1}$. Overall, the number of new measurable components containing $l$ unmatched pairs is 
\begin{align}
    \Delta N_{\rm measurable}^{{N_{\rm CNOT}} = l} =  {6^l {\binom{k}{l}} (10^{k-l}-0.5^{l-1})}
\end{align}

and overall, combining all components with at most $l$ unmatched pairs produces the expression in Eq. \ref{eq:no_meas} of the main text.



\end{document}